\documentclass{aa}
\usepackage{graphicx}
\usepackage{txfonts}
\usepackage{todonotes}
\usepackage[normalem]{ulem}

\newcommand{\gaia}{\textit{Gaia}}
\newcommand\hip{\textsc{Hipparcos}}
\newcommand\tyctwo{\textit{Tycho}-2}
\newcommand{\gbp}{$G_{\rm BP}$}
\newcommand{\grp}{$G_{\rm RP}$}
\newcommand{\xpcol}{\ensuremath{G_{\rm BP}-G_{\rm RP}}}
\newcommand{\phase}[1]{\ensuremath{\approx#1^{\circ}}}
\newcommand{\gband}{$G$--band}
\newcommand{\PP}{\textsf{PhotPipe}}
\newcommand{\calinit}{\textsf{INIT}}
\newcommand{\dsinit}{\calinit\ dataset}
\newcommand{\calonly}{\textsf{CALONLY}}
\newcommand{\dsco}{\calonly\ dataset}
\newcommand{\pdr}{\gaia~DR1}
\newcommand{\gdr}{\gaia~DR2}
\newcommand{\xp}{BP/RP}
\newcommand{\secname}{Sect.}
\newcommand{\equref}[1]{Eq.~\ref{eq:#1}}
\newcommand{\secref}[1]{\secname~\ref{sec:#1}}
\newcommand{\asecref}[1]{Section~\ref{sec:#1}}
\newcommand{\appref}[1]{Appendix~\ref{sec:#1}}
\newcommand{\figref}[1]{Fig.~\ref{fig:#1}}
\newcommand{\afigref}[1]{Figure~\ref{fig:#1}}
\newcommand{\tabref}[1]{Table~\ref{tab:#1}}
\newcommand{\instref}[1]{\inst{\ref{inst:#1}}}
\newcommand{\widefig}[1]{\includegraphics[width=17cm]{#1}}
\newcommand{\colfig}[1]{\resizebox{\hsize}{!}{\includegraphics{#1}}}
\usepackage{amstext}

\begin{document} 

\title{\textit{Gaia} Data Release 2 -- processing of the photometric data}
\titlerunning{Processing of the \gdr\ photometric data}

\author{
  M.~Riello\instref{ioa}              
  \and F.~De~Angeli\instref{ioa}      
  \and D.~W.~Evans\instref{ioa}       
  \and G.~Busso\instref{ioa}          
  \and N.~C.~Hambly\instref{ifa}      
  \and M.~Davidson\instref{ifa}       
  \and P.~W.~Burgess\instref{ioa}     
  \and P.~Montegriffo\instref{oabo}   
  \and P.~J.~Osborne\instref{ioa}     
  \and A.~Kewley\instref{ioa}         
  \and J.~M.~Carrasco\instref{ub}     
  \and C.~Fabricius\instref{ub}       
  \and C.~Jordi\instref{ub}           
  \and C.~Cacciari\instref{oabo}      
  \and F.~van~Leeuwen\instref{ioa}    
  \and G.~Holland\instref{ioa}        
}

\institute{
  Institute of Astronomy, University of Cambridge, Madingley Road, Cambridge CB3 0HA, UK\\ \email{mriello@ast.cam.ac.uk}
  \label{inst:ioa}
  \and
  INAF -- Osservatorio Astronomico di Bologna, via Gobetti 93/3, 40129 Bologna, Italy
  \label{inst:oabo}
  \and
  Institut del Ci\`encies del Cosmos (ICC), Universitat de Barcelona (IEEC-UB), c/ Mart\'{\i} i Franqu\`es, 1, 08028 Barcelona, Spain
  \label{inst:ub}
  \and
  Institute for Astronomy, School of Physics and Astronomy, University of Edinburgh, Royal Observatory, Blackford Hill, Edinburgh, EH9~3HJ, UK
  \label{inst:ifa}
}

\date{Received \textbf{January 26, 2018}; accepted \textbf{February 14, 2018}}

\abstract
    { 
      The second \gaia\ data release is based on 22 months of mission data with an average of 0.9
      billion individual CCD observations per day. A data volume of this size and granularity
      requires a robust and reliable but still flexible system to achieve the demanding accuracy and precision
      constraints that \gaia\ is capable of delivering.
    }
    { 
      We aim to describe the input data, the treatment of blue
photometer/red photometer (\xp)\ low--resolution spectra required to
      produce the integrated \gbp\ and \grp\ fluxes, the process used to establish the internal \gaia\
      photometric system, and finally, the generation of the mean source photometry from the calibrated
      epoch data for \gdr.
    }
    { 
      The internal \gaia\ photometric system was initialised using an iterative process that is
      solely based on \gaia\ data. A set of calibrations was derived for the entire \gdr\ baseline and
      then used to produce the final mean source photometry. The photometric catalogue contains 2.5
      billion sources comprised of three different grades depending on the availability of colour
      information and the procedure used to calibrate them: 1.5 billion gold, 144 million silver, and 0.9
      billion bronze. These figures reflect the results of the photometric processing; the content of the
      data release will be different due to the validation and data quality filters applied during the
      catalogue preparation.
      The photometric processing pipeline, \PP, implements all the processing and calibration
      workflows in terms of Map/Reduce jobs based on the Hadoop platform. This is the first example of
      a processing system for a large astrophysical survey project to make use of these technologies.
    }
    { 
      The improvements in the generation of the integrated \gband\ fluxes, in the attitude modelling,
      in the cross--matching, and and in the identification of spurious detections led to a
      much cleaner input stream for the photometric processing. This, combined with the improvements
      in the definition of the internal photometric system and calibration flow, produced high-quality
      photometry.
      Hadoop proved to be an excellent platform choice for the implementation of \PP\ in terms of
      overall performance, scalability, downtime, and manpower required for operations and maintenance.
    }{}


\maketitle

\section{Introduction}\label{sec:intro}

The European Space Agency \gaia\ mission \citep{GaiaRef} was launched in December 2013. After an
extended commissioning period, science operations began on 25 July 2014. In September 2016,
the first \gaia\ data release \citep[DR1][]{GDR1} was made available to the scientific community, and it
included an astrometric solution based on a combination of \gaia, \hip,\ and \tyctwo\ data \citep{TGAS}
and \gband\ photometry from the first 14 months of operations.

The second \gaia\ data release (DR2) in April 2018 is based on 22 months of mission data and
includes an improved astrometric solution based solely on \gaia\ data (Lindegren at al.\ 2018) and
photometry in \gband, \gbp\ , and \grp\  for approximately 1.5 billion sources. This paper focusses on
the process of calibrating the raw \gband\ photometry and the processing of the low--resolution
spectra to produce and calibrate the \gbp\ and \grp\ photometry. The validation and scientific quality
assessment of the calibrated \gaia\ photometry are discussed in the companion paper, \cite{DWE_DR2}.
We recommend that the \cite{JMC_DR1} paper on the principles of the photometric calibration of the
\gband\ for \pdr\ be read in conjunction with this paper.

The data processing effort for the \gaia\ mission happens in the context of the Data Processing and
Analysis Consortium (DPAC), which is comprised of more than 400 astronomers and software and IT specialists
from over 15 European countries \citep{GaiaRef}. Within DPAC, different groups are set up to handle specific
aspects of the data treatment required to deliver science-ready processed data products to the scientific
community.
The photometric and low--resolution spectra processing system, \PP, consumes a variety of intermediate
data products from other DPAC systems, which, when combined with the low-resolution spectra (see
\secref{instrument}), allows us to derive a consistent set of calibrations that removes most instrumental
effects and establishes a sound internal photometric system that is finally tied to the Vega system
by means of an external calibration process \citep{JMC_DR1}.
A fundamental aspect of the calibration process is that the only stage during which external
(non--\gaia) data are used is in the determination of the external calibration, which uses a set of
well-observed spectro--photometric standard stars (SPSS), see \cite{SPSS}.

For the \gband\ photometry, the processing done by \PP\ does not start from the raw data (i.e.~the
reconstructed satellite telemetry), but from the results of the image parameter determination (IPD),
produced by the intermediate data update (IDU) system, comprising the integrated flux resulting
from a point-spread function (PSF, for 2D observations) or line-spread function (LSF, for 1D
observations) fit of the data, the centroid positions, and relevant statistics and quality metrics.
For the \gbp\ and \grp, \PP\ starts from the raw data and performs all the pre--processing steps
required to produce the uncalibrated integrated flux. Another critical piece of information used by
\PP\ is the cross--match generated by IDU \citep{IDU_XM}: this process identifies transits belonging
to the same astrophysical source after removing spurious detections that are due mostly to artefacts caused by
bright sources.
The pre--processing of the blue photometer/red photometer (\xp)\ spectra involves the bias and proximity electronic module 
non-uniformity mitigation \citep{BIAS}, the straylight \citep{GDR1} mitigation, and the determination
of the geometric calibration mapping the optical distortions and charge-coupled device (CCD) geometry on the focal plane
assembly (FPA). More details on the various pre--processing and the subsequent photometric calibration
process are provided in Sects. \ref{sec:preproc}~and~\ref{sec:calib}, respectively.

The overall processing performed by DPAC is iterative: each data release does not simply include more
data, but it also involves a complete reprocessing from the beginning of the mission, with improved
calibrations and algorithms. In particular, there are a number of significant improvements included in
\gdr. First, the \gband\ pre--processing and IPD have been performed uniformly on the entire data set.
In the first data release, the processing was instead performed on a daily basis by the initial data
treatment \citep[IDT, see][]{IdtRef}: the strict time constraints on IDT and the complexity of the
downlink telemetry scheme meant that it was not always possible to derive and use optimal calibrations
and that data set completeness was not always ensured. A more detailed discussion of the differences
in the IPD process with respect to the \pdr\ can be found in \cite{AGIS_DR2}, \secname~2.
This problem is completely removed in \gdr\ since IDU processes the entire data set in bulk and
therefore can use all the available data to derive the best calibrations (e.g. bias, background, etc.)
to then perform the pre--processing and IPD. Another major improvement with respect to \pdr\ is in the
blacklisting (identification of spurious transits) and cross-match process, which has led to fewer
spurious sources and a cleaner set of transits to work with \citep[see][for more details]{IDU_XM}.
Finally, another important improvement is the handling of micro-meteorites and clanks in the
reconstructed attitude \citep{AGIS_DR2}, which leads to better intra--window source positions. All of
these factors have contributed to a cleaner set of input data
with higher quality for the photometric
processing.

\PP\ features a number of improvements in terms of both
algorithms and processing flow, as we
explain in more detail in \secref{calib}.
Considerable effort has been dedicated over several years to the development of a software system that
is robust and deterministic, but still flexible enough to be able to adapt to the needs of a complex space
mission such as \gaia.
With over 51 billion individual transits, contributing 510 billion individual \gband\ CCD transits and
102 billion low--resolution spectra, achieving high-quality photometry is not only a matter of
devising a sound calibration model \citep{JMC_DR1}, but also of implementing it in a scalable and
resilient fashion.

The processing of \gaia\ data poses several challenges: (1) the intrinsic complexity of the payload
and its operation modes (see \secref{instrument}) leads to a complex data stream in terms of
both raw data and intermediate DPAC data products; (2) the large raw and intermediate data volume (tens of
terabytes) and the fine granularity (0.9 billion individual CCD observations per day) pose demanding
constraints on data storage, I/O, and processing performance; (3) the iterative nature of the DPAC
cyclic processing and of some of the processing algorithms, combined with the
requirement of keeping the overall processing time fixed at each iteration, poses a demanding
requirement on the scalability of the processing systems. Several years before
launch, it became clear that a distributed processing architecture is required to
meet these challenges successfully.
We selected the Hadoop distributed batch-processing system \citep[e.g.][]{Hadoop} for the \PP\
processing architecture. Hadoop provides a reliable distributed file system and a simple
parallelisation abstraction based on the Map/Reduce model \citep[e.g.][]{MapReduce} to develop
distributed data-processing applications. The adoption of Hadoop as the platform for \PP\ has proven
to be very successful in terms of overall performance and robustness, and it is cost effective in terms
of manpower required for operation and maintenance.
\PP\ is the first processing system for a large astrophysical survey project, such as \gaia, to make
use of these technologies. Additional information on the \PP\ processing software and the Map/Reduce
algorithm implementation is provided in \secref{pp}.

Section \ref{sec:instrument} presents a brief overview of the instrument. \asecref{input} provides
a description of the input data used to produce the \gdr\ calibrated photometry.
\asecref{preproc} describes the pre--processing treatment, and \secref{calib} describes the
calibration processing flow. \asecref{pp} presents the architecture developed for the \PP\ processing
pipeline and some aspects of the distributed implementation of the photometric calibration workflow, and
discusses the performance of \gdr\ processing. Finally, some concluding remarks and planned
developments for the near future are given in \secref{theend}. For convenience, a list of the
acronyms used in this paper can be found in \appref{acro}.

\section{Instrument overview}\label{sec:instrument}

\begin{figure}
\begin{center}
\colfig{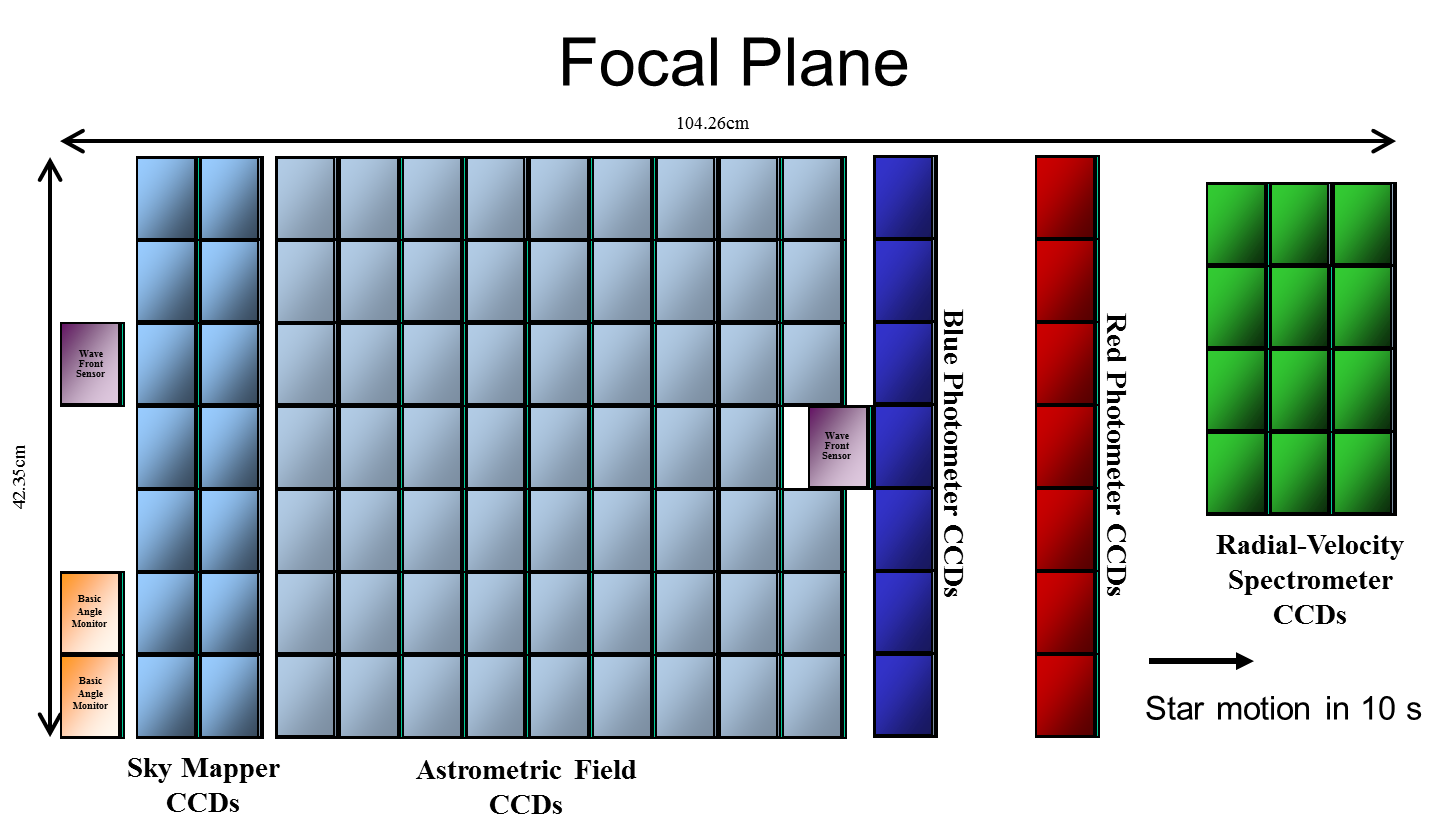}
\caption{
  \gaia\ focal plane, which contains 106 CCDs organised in seven rows. Stellar images travel in the
  along- scan direction from left to right. The 12 CCDs in green are part of the Radial Velocity
  Spectrometer, which will not be described any further because it is not relevant for this paper.
  \label{fig:fpa}
}
\end{center}
\end{figure}

Although a comprehensive description of the \gaia\ mission and payload can be found in \cite{GaiaRef},
in an effort to make this paper more self-contained, this section provides a summary of the mission
and payload aspects that are most relevant to this paper.
The \gaia\ astrometric measurement concept is based on \hip\ and involves two viewing directions
(telescopes) separated by a large fixed angle (the \textup{\textit{\emph{basic angle}}}). The two fields of view (FoV)
are then projected onto a single focal plane. The satellite scans the sky continuously with a fixed
revolution period of six hours. The scanning law designed for \gaia\ provides a full
sky coverage every six months. The sky coverage is not uniform because some areas (nodes) have a very
large number of scans (up to 250 transits per source in the mission nominal length).

{\gaia}'s focal plane uses optical CCDs operated in time-delay
integration
(TDI) mode. Figure~\ref{fig:fpa} shows a schematic view of the focal plane layout.
The CCDs are organised in seven rows. In each row, two CCDs are
dedicated to the sky mappers (SM, one per FoV), nine are dedicated to the astrometric field
(AF, with the exception of row 4, which uses only eight). There are then two CCDs dedicated to the BP and RP. 
The satellite scanning direction is aligned with the rows on the focal plane so that a source image
will enter the focal plane on the SM CCD appropriate for the FoV in which the source is observed and
will then move along the AF CCDs, finally reaching the BP and RP CCDs. A \gaia\ observation of an
astrophysical source is called a \emph{\textit{\textup{FoV transit}}}. The crossing time of a source over an individual
CCD is approximately 4.4 seconds, which therefore provides an upper limit for the exposure time of a
single CCD observation.

All the CCDs in a given row are controlled by a single video processing unit (VPU) that is
responsible for the source detection and confirmation, the definition of the observing mode (see
below), and the recording of all the relevant payload and satellite information that is required for
the ground--based reconstruction process \citep{IdtRef}. The source detection takes place in the SM
CCD for each FoV: if the detection is confirmed by AF1 (i.e. the first of the AF CCDs), the VPU assigns
a window to the source and determines the observing mode for each of the CCDs in the row.
Sources that at detection have an estimated G magnitude of 11.5 or brighter in the SM are
automatically confirmed \citep[see][]{VpuDetection}. Each CCD observation is
acquired by reading a window approximately centred on the source position. The across--scan (AC)
position of the window is computed by the VPU for each CCD, taking into account the estimated AC motion
over the focal plane (i.e. source images do not travel along a straight line on the CCDs).

A complex gating and windowing scheme is implemented on board to control the effective exposure time
and limit the telemetry data volume. Twelve possible gate configurations are available for each
\gaia\ CCD. The gate activation strategy is defined in terms of the onboard detected magnitude and
can be configured independently for each CCD and AC position. In the current configuration, the AF CCD
observations can be acquired without gate or with one of seven different gates.
When activated, a gate will affect all windows that are observed on that CCD during the activation time.
This can create unexpected gated observations for faint sources as well as complex gate situations,
where a gate affects only part of the window.
The window samples can be acquired with or without hardware binning and can be further binned to
reduce the number of bytes required for the downlink. Detections brighter than $G=13$ and $G=11.5$ are
assigned a full-resolution 2D window in AF and \xp,\ respectively. Detections fainter than these limits
are assigned a 1D window (obtained by binning the 2D windows in the AC direction at read--out).
These different configurations are referred to as window--classes.

The \xp\ low--resolution spectra can be acquired either without gate or with one of five active gate
configurations. \xp\ windows are 60 samples long in the along--scan (AL) direction. The spectral
dispersion of the photometric instrument is a function of wavelength and varies in BP from 3 to 27 nm
pixel$^{-1}$ , covering the wavelength range 330–680 nm. In RP, the wavelength range is 630–1050 nm with
a spectral dispersion of 7 to 15 nm pixel$^{-1}$. Because of their larger size, \xp\ windows are more
affected by complex gate cases. Furthermore, contamination and blending issues in dense regions will
particularly affect \xp\ spectra. Due to their larger size, it is generally not possible for the VPU
to allocate a \xp\ window for every detection. Finally, in both AF and \xp\ CCDs, windows can be
truncated in case of overlap. A priority scheme is defined to rule this process.
These non--nominal observations and those affected by a complex gate activation have not been included
in the \gdr\ photometric processing and therefore have not contributed to the released photometry.

\section{Input data}\label{sec:input}

\begin{figure*}
\centering
\widefig{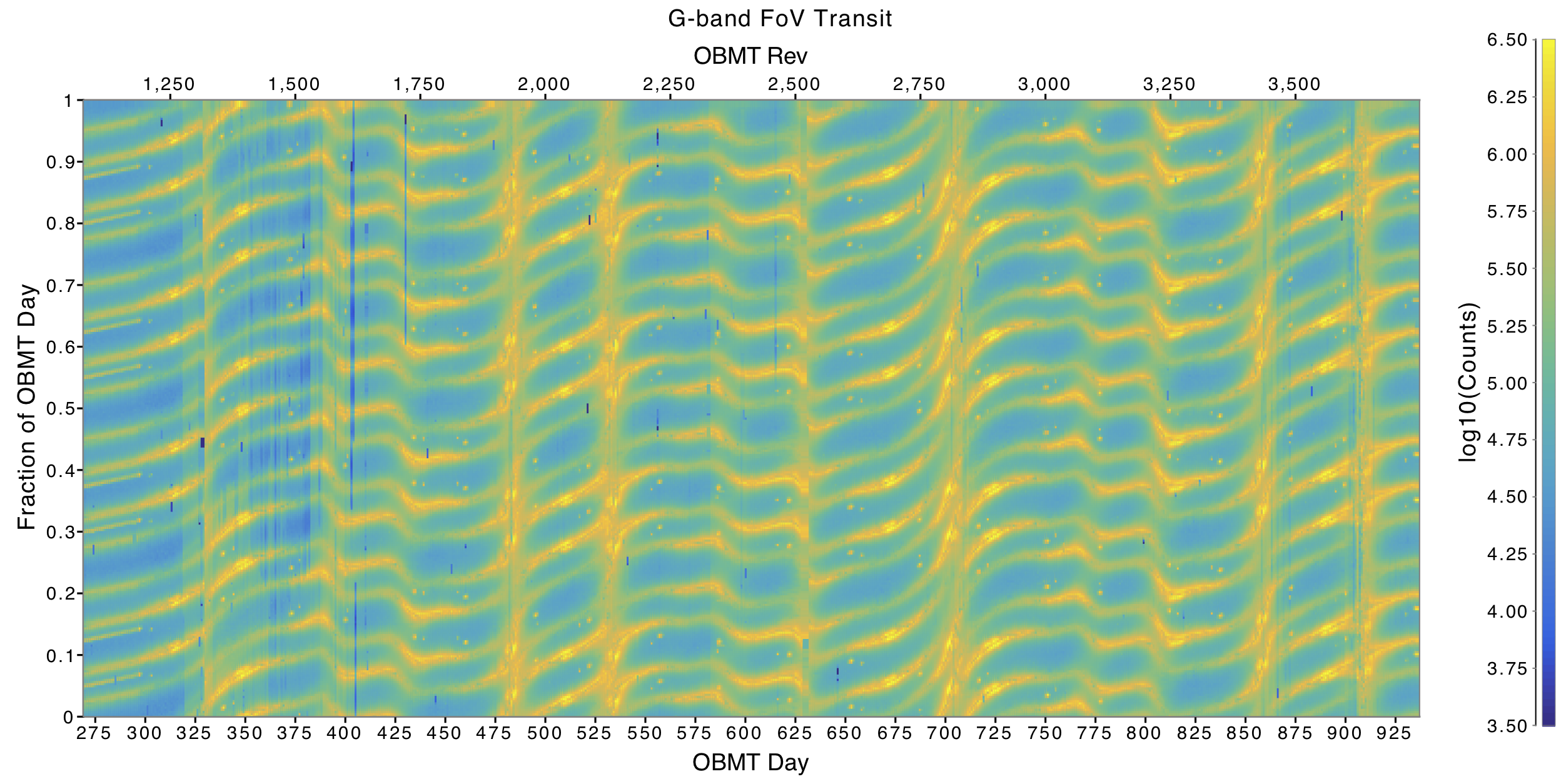}
\caption{
  Temporal density distribution of the $\approx51$ billion \gband\ observations contributing to \gdr.
  Each column in the heatmap shows the density of observations within a given OBMT day for each OBMT
  day.
  The OBMT revolution is shown on the top abscissa axis to facilitate interpretation. The high--density
  features are the Galactic Plane crossing the two FoVs either in the Galaxy inner or outer direction
  (see the text for more details). The gaps related to the events listed in \tabref{events} are also
  visible. Other small gaps are due to telemetry data that could not be included in \gdr\ because
  they were affected by processing problems.
  \label{fig:aohist}
}
\end{figure*}

\gdr\ is based on 22 months of observations starting on 25 July 2014 (10:30 UTC) and ending on 23 May 2016
(11:35 UTC), corresponding to 668 days. When discussing mission events, it is more convenient to use the
onboard mission timeline (OBMT), expressed in units of nominal satellite revolutions (21600 s) from an
arbitrary origin. An approximate relation to convert beween OBMT revolutions and barycentric coordinate
time (TCB) is provided by Eq. 3 in \cite{GaiaRef}. Hereafter we use \emph{rev} to mean OBMT revolutions.
The period covered by \gdr\ extends from 1078.38 to 3750.56 rev.

There are a number of events that have to be taken into account in the photometric
calibration for \gdr: two decontamination and two re-focussing events (see \tabref{events}).
Decontaminations are required to mitigate the throughput loss caused by water-based contaminant present
in the payload \citep{GaiaRef}. The size of the systematic effect due to the contamination is orders
of magnitude larger than the expected level of any systematic effect.
Decontamination campaigns are required to recover the optimal performance: they involve actively
heating the focal plane and/or some of the mirrors to sublimate the contaminant. Refocussing events
have been carried out after each decontamination event.

\begin{table}
\begin{center}
\begin{tabular}{lccc}
\hline
\multicolumn{1}{c}{Event} & \multicolumn{2}{c}{OBMT range [rev]} & \multicolumn{1}{c}{Duration}\\
\multicolumn{1}{c}{} & \multicolumn{1}{c}{Start} & \multicolumn{1}{c}{Stop} & \multicolumn{1}{c}{[rev]}\\
\hline
  Decontamination & 1316.492 & 1324.101 & 7.609\\
  Refocussing     & 1443.950 & 1443.975 & 0.025\\
  Decontamination & 2330.616 & 2338.962 & 8.346\\
  Refocussing     & 2574.644 & 2574.728 & 0.084\\
\hline
\end{tabular}
\end{center}
\caption{\gdr\ mission events relevant for the photometric calibration process.\label{tab:events}}
\end{table}

It should be noted that after a decontamination event is completed (i.e. the active heating is turned
off), a much longer time is required for the focal plane to return to its nominal operating
temperature. Although science operations resume at the end of a decontamination campaign, the data
quality will not be nominal until thermal equilibrium has been reached. Data obtained during
the time ranges listed in \tabref{events} have not been included in the photometric processing.
Furthermore, these events create discontinuities in the instrumental behaviour that can be used as
natural breaking points for the definition of the photometric calibrations (see \secref{calib} for
more information).

\afigref{aohist} shows the density of \gband\ FoV transits observed by \gaia\ in the time range
covered by \gdr\ (abscissa) with intra--day resolution (ordinate). For a given abscissa
position (i.e. one OBMT day), the ordinate shows the density variation within the four OBMT
revolutions of that day, thus allowing a much higher level of detail to be visible compared to a
standard histogram. Several features are visible, in particular:

\begin{itemize}
\item
  Sixteen daily Galactic Plane (GP) crossings: eight in the inner and eight in the outer direction of the
  galaxy, four for each FoV. The GP features become progressively steeper in the plot
  because the spacecraft spin axis becomes perpendicular to the GP itself thus leading to a
  GP scan (GPS) when the two \gaia\ FoVs effectively scan the GP continuously for
  several days (e.g. at $\approx1945$ rev and then again at $\approx2120$ rev, etc.)

\item
  The decontamination events (see \tabref{events}), which manifest as gaps in the data. Refocussing
  events are harder to spot because their duration is much shorter.

\item
  Outages in the daily processing pipelines, which manifest as minor gaps. These outages meant that
  some satellite telemetry was not actually available for \gdr\ processing, but will disappear in
  future date release. Other gaps are instead caused by genuine spacecraft events and will never
  disappear.

\item
  The eight thin streaks visible before 1200 rev are due to the LMC crossing the two FoVs at each
  revolution during the ecliptic poles scanning mode (see below). After this, the LMC is still
  visible as increased density spots at periodic intervals.
\end{itemize}

From the start of scientific operations up to 1185.325 rev, \gaia\ observed following the ecliptic
poles scanning law (EPSL), which meant that both FoVs were scanning through the north and south
ecliptic pole at each revolution (with the scanning direction changing at the same rate as the Sun)
and the spin axis moving along the ecliptic at a rate of $\approx1^{\circ}$ per day. The main aim of
this scanning mode was to provide end-of-mission coverage for a limited portion of the sky in a very
short amount of time for the purpose of bootstrapping the photometric calibrations and to assess the
scientific performance of the mission \citep[see \secname~5.2 in][for more information]{GaiaRef}.
Unfortunately, the period leading up to the first decontamination proved to be very unsuitable for
the purpose of establishing the photometric system as originally planned because of the high level of
contamination, and more importantly, because of its strong temporal variation. In \secref{calib} we discuss
the implications of this for the flux calibration process. After the EPSL phase, the scanning law was
transitioned to the nominal one.

The main inputs to \PP\ are 1) the IDU pre--processed \gband\ transits, providing centroid, IPD
information, and basic acquisition and quality information; 2) the IDU cross--match associating each
transit to a source; 3) the source astrometry and reconstructed spacecraft attitude produced by the
astrometric global iterative solution \citep[AGIS,][]{AGIS_REF} system; and 4) the raw \xp\
low--resolution spectra.
The IDU PSF (for 2D observations) and LSF (for 1D observations) used for \gdr\ are similar to those
used for \pdr\ that are described in \cite{IdtRef}.
It is worth mentioning that the PSF/LSF models have been derived from mission data in the range
3350--3365 rev and do not model the colour and time dependencies. This can create systematic
effects on the derived fluxes that are time and colour dependent due to the time--varying
contamination: these systematics can become more noticeable when handling epoch data, but are
less critical for the source photometry, where they will result in increased errors of individual
sources.

The \gaia\ onboard detection algorithm \citep{VpuDetection} operates without a source catalogue,
which means that the spacecraft telemetry provides only transit--based information: no further
knowledge about the association of each transit to a given astrophysical source is available.
Associating transits to individual sources is the main goal of the cross--match task performed by IDU
\citep{IDU_XM}.
A pre--processing stage identifies spurious detections that are
due to artefacts caused by bright sources or
extended objects. The cross--match process then associates each individual transit with a source.
Although the process reuses the source identifiers that have been created in previous iterations (e.g.
\pdr\ in this case), it should be noted that the source identifier is simply a label: what actually
provides an identity to a source are the transits that are associated with it since that will eventually
determine the source astrometric parameters and photometry. For this reason, it is not possible to
directly compare individual sources between different \gaia\ releases, and \gdr\ should be treated as a
new and independent catalogue.
\tabref{counts} summarises the number of input records processed by the \PP\ system: these represent a
superset of the \gdr\ content as low--quality, incomplete, and/or non--nominal data have been excluded
from the release \citep[see][]{CU9}.

\begin{table}
\begin{center}  
\begin{tabular}{lr}
\hline
Type & \multicolumn{1}{c}{No.~records}\\
\hline
\gband\ FoV transit          & 51,712,381,972\\
\xp\ raw FoV transit         & 51,715,475,265\\
Spurious detections          & 10,737,486,581\\
IDU cross--match sources     &  2,582,614,429\\
AGIS sources                 &  2,499,375,298\\
\hline
\end{tabular}
\end{center}
\caption{
  Summary of the main input records used by \PP\ for \gdr. The spurious detections are a subset of
  input \gband\ transits and have been excluded from the cross--match because they may be associated
  with artefacts from bright sources. AGIS sources refer to the number of source records with sky
  positions determined by AGIS.
  \label{tab:counts}
}
\end{table}

\section{Pre-processing}\label{sec:preproc}

As mentioned in the previous section, the raw SM and AF CCD transits are first processed in IDU,
which takes care of bias correction, background determination, and
removal, and estimates centroid positions and the \gband\ (uncalibrated) flux based on PSF/LSF fitting
\citep[see][]{IdtRef}. The pre--processing for the \xp\ CCD transits is carried out by \PP\ instead and
is described in this section.

The pre--processing stage is required to prepare the raw integrated epoch fluxes in all bands (at the
CCD level) for the calibration step. For all CCD transits, we compute the predicted positions of the
image centroid on the CCD from the reconstructed satellite attitude, the geometric calibration
(expressed as a correction to the nominal field angles, as described in \citealt[][Sect.~3.4]{AGIS_REF}),
and the source astrometric parameters as derived by AGIS \citep{AGIS_DR2}: this is essentially the
inverse of the operation described in \citet[][Sect.~6.4]{IdtRef}. Since the AC centroid position of
the image is only available for the 2D \gband\ transits, which are $\approx1\%$ of the total, the
flux calibration models (see \secref{calib}) use the predicted AC position as the best available
information on the AC location of the source image on the CCD. In \pdr, the predicted AC position could
not be computed consistently for all transits, and the calibration models therefore used the AC window
centre, which is equivalent to assuming that each source image is perfectly centred in the AC
direction.
The \xp\ integrated fluxes and spectrum shape coefficients \citep[SSCs, see][]{JMC_DR1} are produced
by \PP\ from the \xp\ spectra after they have undergone several pre--processing steps: correction and
mitigation of electronic offset effects (\secref{debias}), background correction (\secref{straylight}),
and AL geometric calibration (\secref{geocal}).

\subsection{Correction and mitigation of electronic offset effects}\label{sec:debias}

\begin{figure}
\centerline{\includegraphics[scale=0.2,clip=true,trim=0 25 0 25]{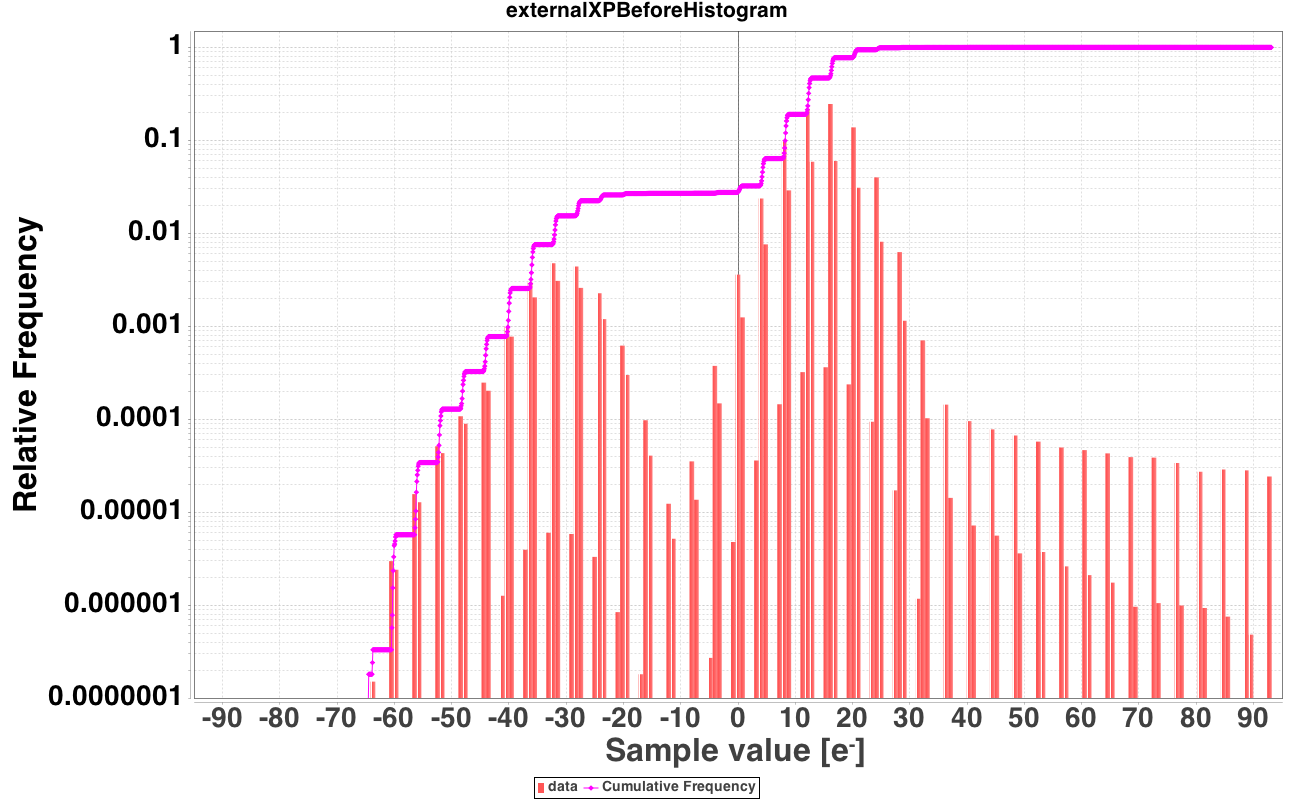}}
\centerline{\includegraphics[scale=0.2,clip=true,trim=0 25 0 25]{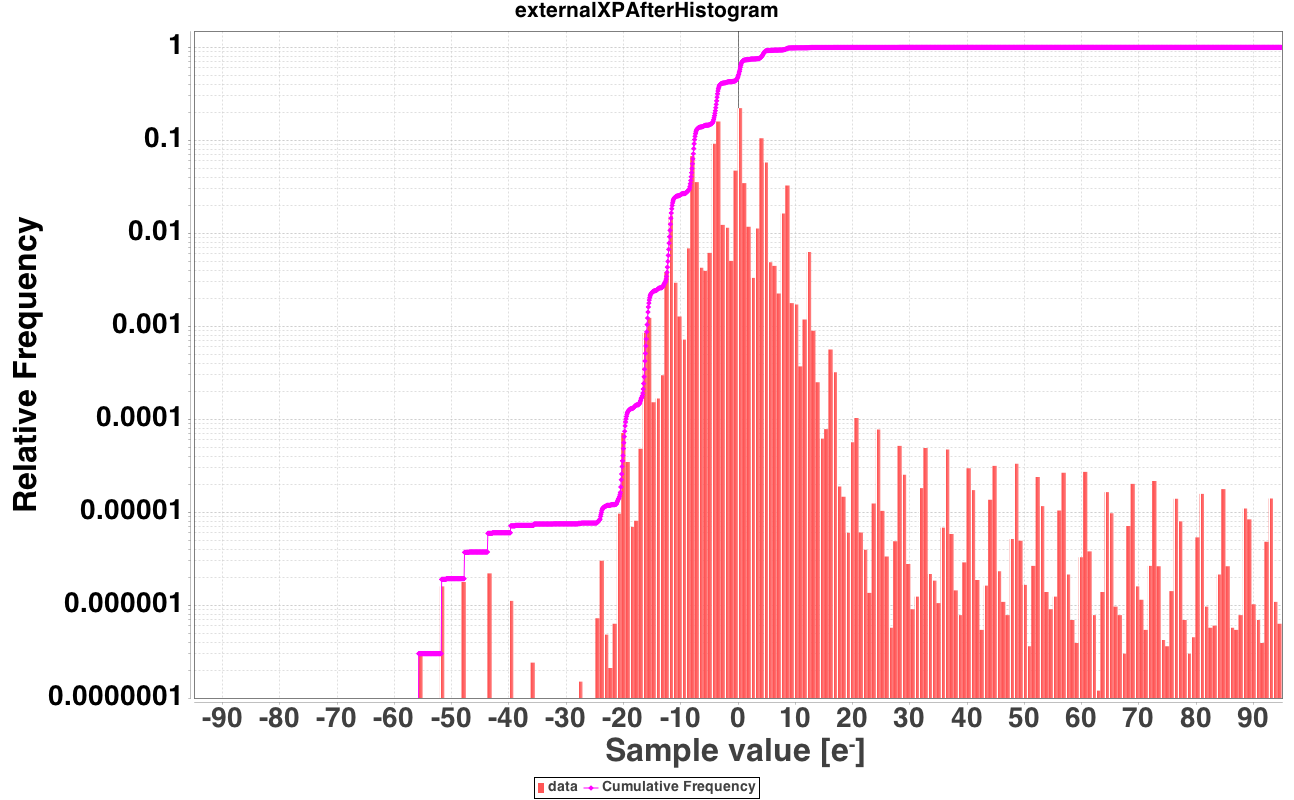}}
\caption{
  Sample distribution in empty windows affected by gate~5 activation (and hence limited in integration
  time to~32~ms) when they are bias corrected by scalar prescan level only (top) and when they are bias
  corrected using the full bias instability model (bottom). The sample distribution in the
  latter has a near--normal distribution (disregarding the negligible few outliers on the positive
  side of the distribution resulting from stray photoelectric flux and prompt--particle events, etc.)
  with Gaussian equivalent $\sigma=5.2e^-$ and is dominated by the video chain read-noise-limited
  performance for that device (see \citealt{BIAS} for further details). In both panels the magenta
  line shows the cumulative distribution.
  \label{fig:bppemnu}
}
\end{figure}

The electronic zero-point offset on the CCD amplification stage (commonly referred to as the bias
level) is in principle separable from nearly all other calibrations. However, the complexity of the
\gaia\ CCD design and operation leads to quasi--stable behaviour that in turn considerably complicates
the determination of the additive correction to be applied to the data at the beginning of the
processing chain \citep{BIAS}. In addition to the normal zero point of the digitised sample data
(which in the case of \gaia\ is measured via periodic prescan samples), offset excursions
are present on any given data with amplitudes of up to~$\approx16$~ADU~($\approx64e^-$) in~\xp\ depending on
the timing of that sample in the serial scan and on the number (if any) of fast--flushed pixels
preceding the sample. Furthermore, the onset of the excursions and recovery as normal samples are
read is a non--trivial function of the flushing, reading, and occasional pausing in the serial scan
(exhaustive detail is given in \citealt{BIAS}).
Hence the full mitigation of these electronic effects involves effectively reconstructing the readout
history of the CCD in a window of $\approx30$ s centred on each detection. For \gdr, all the
required calibrations are determined in the IDT and in the First Look CCD one-day calibration subsystems,
but the process of determining the correct bias level for each sample still requires readout
reconstruction from observation log records that are telemetered as part of the auxiliary data
streams into the on--ground processing pipelines. 

\afigref{bppemnu} shows an example of the effectiveness of the offset instability correction
procedure for the~BP~CCD in row~3 of the \gaia\ focal plane. This device shows the largest excursions
from the gross electronic zero point amongst the astro--photometric devices. For this illustration we
have chosen samples that have been affected by a gate~5 activation. We note that these are not samples
from windows containing objects that have triggered a gate~5 activation; we have chosen instead
samples from empty windows (also known as `virtual objects', VO) 
that are observed at the same time as such a window containing a very bright star, but at different AC
positions within the same CCD. The integration time of these samples is limited to~32~ms, resulting in
very small photoelectric background correction and hence no possibility of significant residual
systematic errors from that correction, accurate calibration of which also depends on bias
correction, of course. These selected samples are the closest approximation we can achieve to `dark' observations in \gaia,\
which scans continuously with no shutter. The distribution of sample values corrected for prescan level
only shows a high--amplitude systematic residual pattern that
is introduced by the offset instability
excursions resulting from the multifarious sample serial timings as the observed windows transit the
CCD. 
The core distribution of samples corrected with the full bias model is, however, limited to a
near--normal distribution equivalent to the distribution expected given the video chain detection noise
performance. We note that a relatively small number of samples remain uncorrected in this example, which
features data from 1973--2297 rev. These arise in \gdr\ in situations where the data stream is
incomplete and on--ground readout reconstruction is consequently inaccurate. This problem affected
\pdr, is significantly reduced in DR2, and will be further reduced in DR3.

\begin{figure}
\begin{center}
\colfig{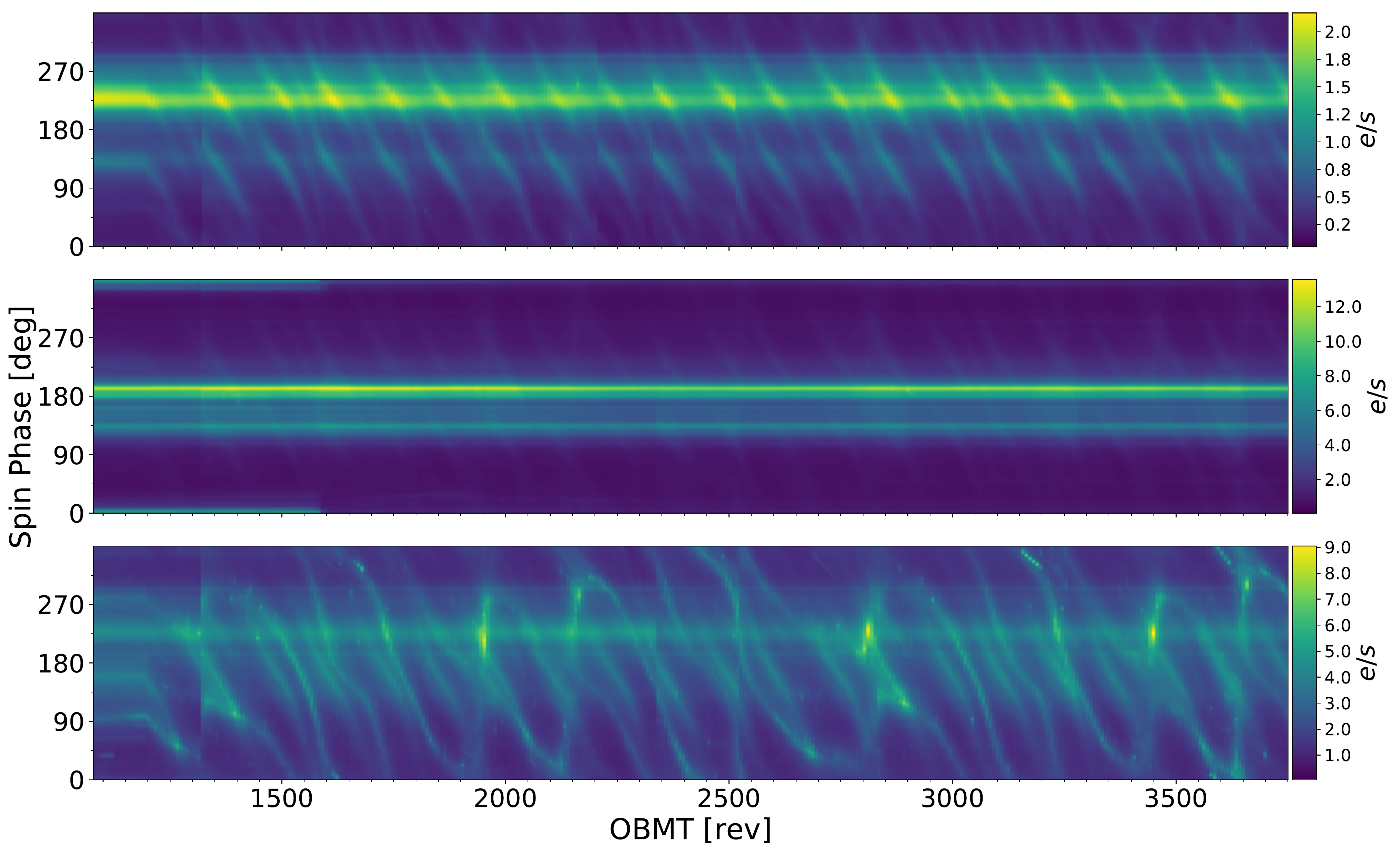}
\caption{
  Straylight background level evolution in BP. Each panel show the straylight level as a function
  of time and satellite spin phase. The top panel shows BP row 1, which has the lowest overall
  straylight level. The central panel shows BP  row 7, which has the highest straylight level, but
  shows a very stable pattern. The bottom panel shows BP row 5, which has a higher and extremely
  variable straylight level. See the text for further discussion.
  \label{fig:slevol}
}
\end{center}
\end{figure}

\subsection{\xp\ straylight mitigation}\label{sec:straylight}

As has been reported for \pdr, the large--scale background has a major contribution in all instruments
from light scattered by loose fibres on the solar shield edges that enter the FPA via illegal optical
paths \citep[][\secname~5.1.3]{IdtRef}. However, cosmic sources also contribute significantly. The background
mitigation for the G-band is performed in IDU and involves fitting a 2D spline as a function of time and
AC position for each CCD. The variation of the straylight with time in SM/AF is therefore captured reasonably
well.

In both \pdr\ and DR2, the \xp\ background mitigation performed by \PP\ primarily involves the
determination of the straylight component. The straylight pattern depends on the spin phase of the
satellite and is stable over several tens of revolutions. Instead of explicitly modelling the time
dependence of the straylight pattern for each CCD, an independent solution is determined on every
set of consecutive $\approx8$ rev time intervals in the \gdr\ dataset (excluding the events in
\tabref{events}). The calibration uses the VO empty windows, which are allocated by each VPU
according to a predefined spatial and temporal pattern.
For each VO, \PP\ determines the median level and constructs the AC versus spin-phase straylight map by
taking the median level from all contributions in each AC/phase bin. For \gdr, the resolution of the
maps was $\approx100$ pixels in the AC directions (20 bins AC) and $1^{\circ}$ in the phase direction
(360 bins in phase).

The resulting maps can occasionally contain gaps (i.e. empty bins) caused by missing data or gaps in
the reconstructed attitude. To reduce the impact of gaps, the maps are processed on the fly to fill
the gaps via interpolation. The process first attempts to fill the gaps by interpolating along the
phase and then fills any remaining empty bins by interpolating along the AC dimension. In both cases, we
used linear interpolation by searching the nearest non--empty bin within a configurable range (four bins
for phase and three bins for AC). Even after this interpolation process, it is possible for empty bins to be
present in the case of very large gaps. These bins are assigned a default value equal to their nearest
phase bin and are flagged to ensure that they will not be used by the background level estimation
process\footnote{
  This stage is required because empty bins will otherwise confuse the bicubic interpolator used
  for the straylight level estimation process.
}.
The straylight level to be removed from each transit is then determined from the appropriate
pre--processed map via bicubic interpolation.

One effective way to visually evaluate the stability of the straylight background over time is to
create an animation from the individual straylight maps. An alternative approach is shown in
\figref{slevol}, where we generate an average straylight profile (level versus spin phase) from each map
and then display all the profiles as a function of OBMT revolution. Three cases are shown to
illustrate the challenges faced in mitigating the straylight background. The top panel shows BP row 1:
the CCD least affected by straylight, peaking at nearly 3 $e^-/s$, the main peak is located at a phase
of $\phase{225}$ , and its location appears to be quite stable over time. A secondary, fainter feature
is visible at a phase of $\phase{130}$. Both features are very stable during the EPSL period, while
they appear to progressively drift in phase when the satellite follows the nominal scanning law.
The central panel of \figref{slevol} shows the straylight evolution for BP row 7: the CCD with the
highest level of straylight. Although the level is much higher than in row 1 (see the different range
covered by the colour scale in the corresponding plots), the pattern is very stable for all of the
three main features: the brightest at phase $\phase{190}$ , and two fainter features at phase $\phase{135}$
and $\phase{170}$.
Finally, the bottom panel shows BP row 5, where the overall straighlight level is not as strong as in
row 7, but shows a very complex temporal evolution. There are four peaks at phases $\phase{90}$,
$\phase{150}$, $\phase{225}$ , and $\phase{280}$ that are stable during the EPSL period, but then appear
to drift along the entire phase range in a cyclic fashion, creating a complex pattern. The peak at
$\phase{225}$ also appears to maintain a stable component over the entire time range.

\begin{figure*}
\centering
\resizebox{0.44\hsize}{!}{\includegraphics{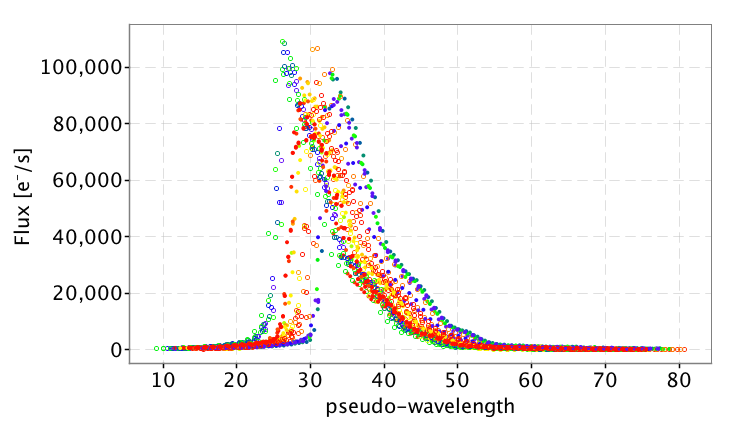}}
\resizebox{0.51\hsize}{!}{\includegraphics{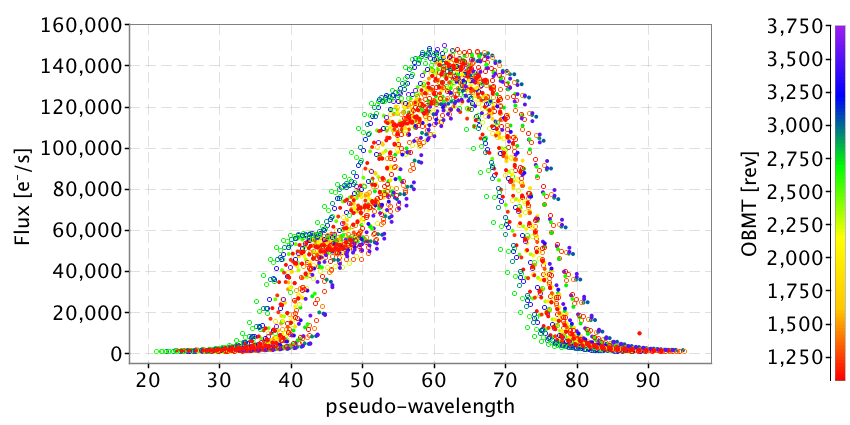}}
\resizebox{0.44\hsize}{!}{\includegraphics{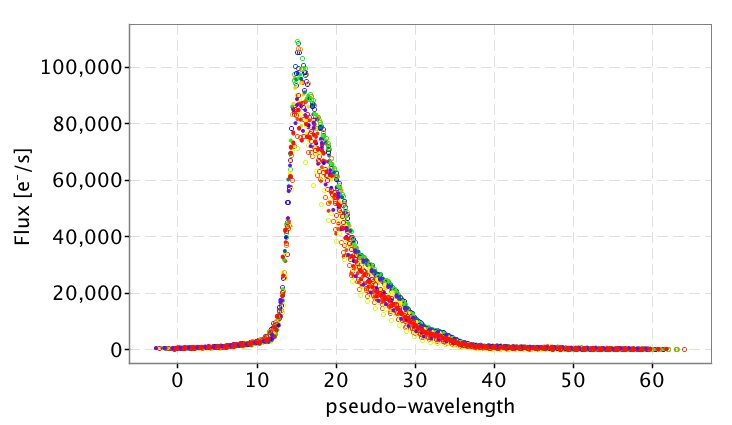}}
\resizebox{0.51\hsize}{!}{\includegraphics{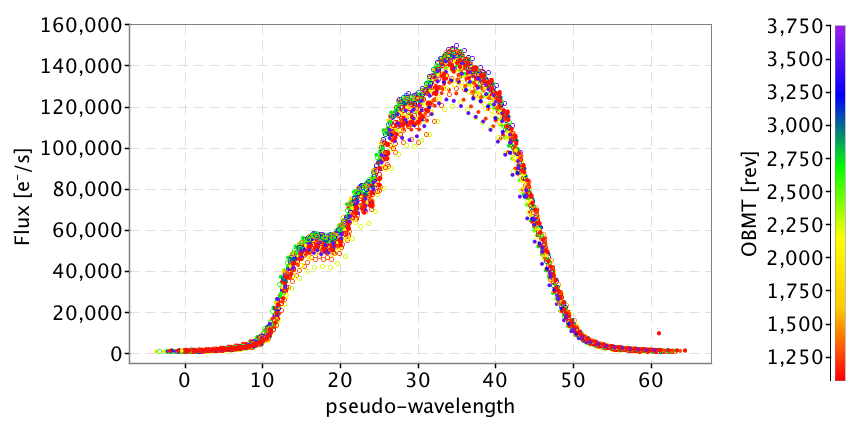}}
\caption{
  Epoch spectra for one of the SPSS sources. BP and RP are shown in the left and right panels,
  respectively. The {\em top row} shows the epoch spectra aligned using the nominal geometric
  calibration only. The {\em bottom row} shows the same epoch spectra after application of the
  differential geometric calibration computed by \PP. Filled symbols are used for the preceding FoV,
  while open symbols show the following FoV. Symbols are colour--coded by time in OBMT-rev as
  indicated by the colour bar. 
  \label{fig:spssepochs}}
\end{figure*}

\subsection{\xp\ AL geometric calibration}\label{sec:geocal}

Although low--resolution spectral data are not part of \gdr, some aspects of the \xp\
spectral processing are very important in the generation of the photometric catalogue and should
therefore be described in this paper.
As in the case of \gband\ observations, spectra are also collected in small windows centred around the
detected sources. The incoming light is dispersed in the AL direction by a prism. However, the flux at
each wavelength is additionally spread over a range of samples according to the LSF appropriate for
that wavelength. This means that the flux collected in each sample will have contributions from a
range of wavelengths whose width depends on the FWHM of the LSF at the various wavelengths. The
size of these contributions depends on the source spectral energy distribution.

Several calibrations are required to convert the acquired flux per sample into the flux in a  band
covering a specific wavelength range. While for the generation of integrated \gbp\ and \grp\ a simple
sum of the flux in all the samples of a window is sufficient, an accurate calibration of the AL
coordinate within the window, in terms of absolute wavelength, is required to extract more detailed
colour information in the form of SSCs \citep[see][and Appendix \ref{sec:sscs}]{JMC_DR1}. 
The dispersion calibration provides a relation between the AL coordinate within the window and the
absolute wavelength scale. The nominal pre--launch dispersion calibration has been adopted for \gdr.
This was derived from chief--ray analysis from fitting a polynomial function to the unperturbed
EADS--Astrium Gaia optical design with a maximum fit uncertainty of 0.01 AL pixel \citep{PLM00108}. 
However, the polynomial function is defined with respect to the location within the window of a
specific reference wavelength, chosen to be well centred in the instrument wavelength range, and
therefore its application is complicated by the fact that sources are often not well centred (due to
inaccuracies in onboard detection window assignment).
The location of the source centroid within the window can be predicted using our knowledge of the
source astrophysical coordinates, the satellite attitude, and the layout of the CCDs in the focal
plane, i.e. their geometry.
Astrophysical coordinates and satellite attitude are best calibrated using the \gband\ data in the
AGIS system \citep{AGIS_REF}, while the geometry of the \xp\ CCDs is calibrated as part of the \PP\
pre--processing.

The AL geometric calibration is computed differentially with respect to the expected nominal geometry
based on pre--launch knowledge of the CCD layout. An initial guess of the source location within the
window is obtained by adopting the nominal geometry. The calibration process aims at modelling the
corrections to be applied to the nominal predicted positions. For more details on the calibration
procedure and on the model definition, see \cite{JMC_DR1}.

\begin{figure}[h!t]
\begin{center}
\colfig{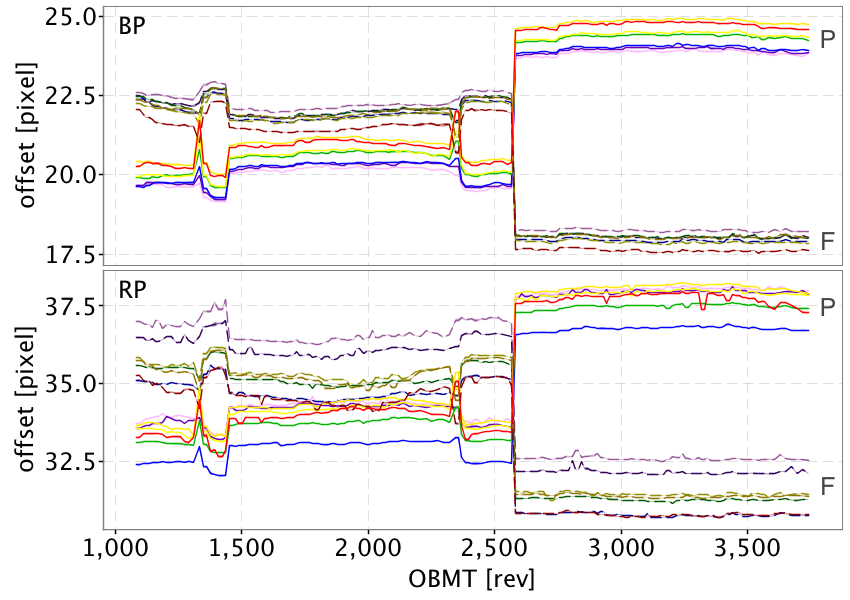}
\caption{
  Evolution in time of the geometric calibration, relative to the nominal geometry, evaluated at 
  the centre of the CCD in the across-scan direction and for different CCDs (BP in the top panel,
  RP in the bottom panel; rows from 1 to 7 are shown in red, orange, yellow, green, blue, purple, and 
  pink). The preceding FoV is shown with brighter
  colours and solid lines, while darker shades and dashed lines are used for the following FoV, as 
  indicated by the labels P and F in the plot area.
  \label{fig:geocal}
}
\end{center}
\end{figure}

\afigref{spssepochs} shows the epoch spectra available for one of the SPSS \citep{SPSS} used in the
external calibration. This source was chosen because it is quite bright and has a large and
well--distributed number of epochs.
The top row shows the epochs calibrated using only our nominal knowledge of the CCD geometry (BP
spectra are shown on the left and RP spectra are shown on the right). The bottom row shows the same
epochs after the application of the calibration produced by \PP.
In all panels of \figref{spssepochs}, the location of each sample and the corresponding flux have
been shifted and scaled respectively according to the differential dispersion across the focal plane.
This creates an internal reference system that is referred to as a pseudo--wavelength scale.
\afigref{geocal} shows the actual calibration evaluated at different locations (CCD edge cases in the
AC direction are shown with dashed and dotted lines, while the value in the centre of the CCD is shown
with a solid line) on the various CCDs (in different colours, red corresponding to row 1 and purple to
row 7) for BP in the top panel and RP in the bottom panel at various times in the period covered by
\gdr. 

When comparing Figs. \ref{fig:spssepochs} and \ref{fig:geocal}, it is easy to see that the
calibrations can reproduce the significant offsets observed for approximately simultaneous spectra
from different FoVs. For instance, epochs in the period 3000--5000 rev (colour-coded in blue in
\figref{spssepochs}) show an offset between the two FoVs of several AL pixels and a wide separation in
the calibrations for the two FoVs. The difference between the two FoVs is much smaller in other
periods, hardly noticeable, for instance, in the period 1750--2000 rev (colour--coded in yellow in
\figref{spssepochs}), which is confirmed by the calibrations evaluated in the same period.
Discontinuities in the calibrations shown in \figref{geocal} are clearly related to decontamination
and refocus activities (see \tabref{events}). In general, the RP calibration is noisier because the features
in the RP spectrum are smoother than those in the BP spectrum.
The standard deviation of the solution evaluated in the period 2700--3700 rev is 0.05 for BP and 0.12
for RP in AL pixels. These are equivalent to 0.4 nm and 1.3 nm, respectively, in the central part of the
spectrum. Uncertainties of this size are negligible when computing the spectrum shape coefficients used
for the photometric calibrations (for more details, see \appref{sscs}). 

Systematic errors in the geometric calibration parameters would not affect the photometric
calibrations as they will simply result in a slightly different set of SSC bands being used for the
definition of the colour information.
After dispersion and geometry calibrations have been carried out, it becomes possible to estimate the
flux in given passbands, such as those used to define the SSCs. We call these `raw' or
`uncalibrated' SSCs, even though the calibration process described above has been applied in order to
generate them. The fluxes obtained at this stage will still be affected by differences in the CCD
response and in the LSF across the \xp\ strips of CCDs. None
of the spectra shown in this paper are calibrated for response and LSF. These effects are calibrated out
in the internal calibration step.
For \gdr,\ the uncalibrated SSCs and integrated \gbp\ and \grp\ are calibrated following the same
procedure as applied to the \gband\ and is described in \secref{calib}.

\section{Calibration of \gaia\ integrated photometry}\label{sec:calib}

\subsection{Overview}\label{sec:calsum}

\begin{figure}
\begin{center}
\colfig{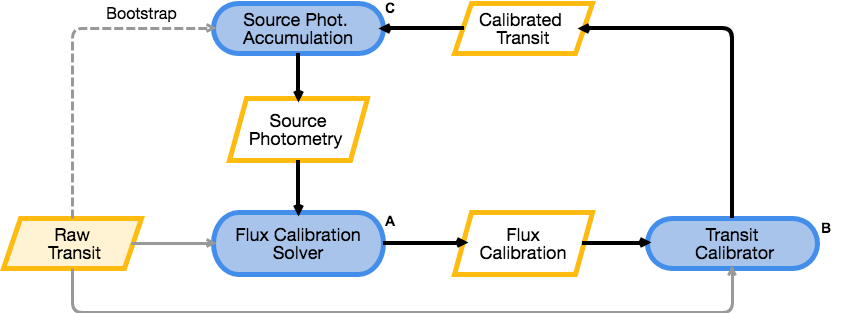}
\caption{
  Iterative internal calibration flowchart. The process is started by bootstrapping the reference
  source photometry from the raw transits, and it then proceeds by iteratively deriving new flux
  calibrations, which are then used to produce an updated set of reference fluxes. The calibration loop
  is represented by the three processes labelled A, B, and C; the dataflow is represented by thicker
  arrows. See the text for additional information.
  \label{fig:calloop}
}
\end{center}
\end{figure}

The calibration of the \gband\ and \xp\ integrated photometry is based on the principle of first
performing a self--calibration on an internal system using only \gaia\ data, followed by an external
calibration to link the internal to the external system \citep{JMC_DR1}. The internal calibration
workflow is illustrated in \figref{calloop} and involves establishing the internal \gaia\ photometric
system as defined by a set of standard sources with defined reference \gband\ and \xp\ integrated
fluxes.
These standards are then used to derive the set of photometric calibrations required to calibrate all
individual epochs. These calibrated epochs are then combined to derive the source photometry in the
internal photometric system. Since the reference fluxes for the standard sources are not known a
priori, they are derived via a simple iterative bootstrap procedure from the uncalibrated source
photometry. Each step is described in greater detail below:
\begin{enumerate}
\item
  Compute the raw source photometry from uncalibrated epochs to use as starting values for the
  reference fluxes. This step is represented in Fig.~\ref{fig:calloop} by the process labelled C,
  operating on raw transit input as represented by the thin dashed line labelled `bootstrap'.
\item
  Derive a set of calibrations based on the current set of reference fluxes. This step is represented
  in \figref{calloop} by the process labelled A, operating on the raw transits and the reference
  fluxes for the corresponding sources. The calibration model is described in more detail in
  \cite{JMC_DR1}.
\item
  Apply the calibration to the individual epochs. This step is represented in \figref{calloop} by the
  process labelled B, operating on the raw transits and using the calibration derived at the
  previous step to produce the calibrated epoch photometry.
\item
  Recompute the source photometry to provide an updated set of reference fluxes. This step is
  labelled C; it is the same as the first step, but operating this time on calibrated transits
  instead of raw ones.
\item
  Iterate by repeating steps from two to four until convergence is reached. This is the calibration
  loop that is shown in \figref{calloop} as the three processes labelled A, B, and C; the dataflow
  is represented by thicker lines.
\end{enumerate}

The calibration model is composed of a large--scale (LS) and a small--scale (SS) component. The LS
component tracks the fast changes in the instrument over timescales of a few revolutions ($\approx  $
one day), whereas the SS component tracks more stable sensitivity variations and effectively provides a
1D flat--field equivalent.
Instead of explicitly modelling the time dependence in the LS calibrations, we simply computed a number
of independent solutions on $\approx4$ rev time ranges spanning the \gdr\ dataset, but excluding the
decontamination and refocussing events listed in \tabref{events}.
The LS model used for \gdr\ features a quadratic dependency on the AC position of the transit and a
linear dependency from the source colour. The colour information is expressed in terms of the spectral
shape coefficients (SSC), which are derived by integrating the \xp\ spectra on a predefined set of
top--hat bands providing four integrated fluxes from the BP spectrum and four from the RP spectrum
\citep[see \secname~4 in][]{JMC_DR1}. The main advantage of using SSC--based colours is that it
allows the use of lower order dependencies in the calibration models than when using a plain (e.g.
$\xpcol$) colour by providing more detailed colour information. However, for the SS calibration, the
\gdr\ model simply involves a zero point.

\subsection{Robustness}

When computing the least-squares (LSQ) solutions for the LS and SS models, we exclude from the
solutions non--nominal observations, that is, observations that are either truncated or that have been
acquired with a complex gate configuration. In addition, we filter observations that have been flagged
as problematic in the acquisition or IPD processes: these observations are tagged with the
corresponding set of problems and are then used to generate a report, attached to each individual
solution, describing how many observation have been excluded and for which reasons.
One additional filter is used to exclude outliers that might originate from cross-match problems by
excluding any observation exhibiting a difference between the transit flux and the source reference
flux larger than 1 magnitude. These filters are only applied when solving for the calibrations. A
different robustness process handles the rejection of unsuitable epochs when generating the calibrated
source photometry, as described in \secref{srcphot}.

Each LSQ solution is computed iteratively: at a given iteration, we use the solution computed at the
previous iteration to reject observations that are discrepant by more than $N\sigma$. At each iteration, the
rejection process will evaluate the residuals of all measurements (including those that were rejected
in a previous iteration). In \gdr,\ the rejection process has been configured with a $5\sigma$
rejection threshold and a maximum number of ten iterations. The rejection process is attempted only if
there are at least 20 observations contributing to the solution. This approach requires all available
observations to be kept in memory during the iteration process: since the calibration models have a
low number of parameters, this is never a problem, even when there are millions of observations
contributing to a given calibration solution.

\subsection{Time--link calibration}\label{sec:tlc}

\begin{figure*}
\centering
\widefig{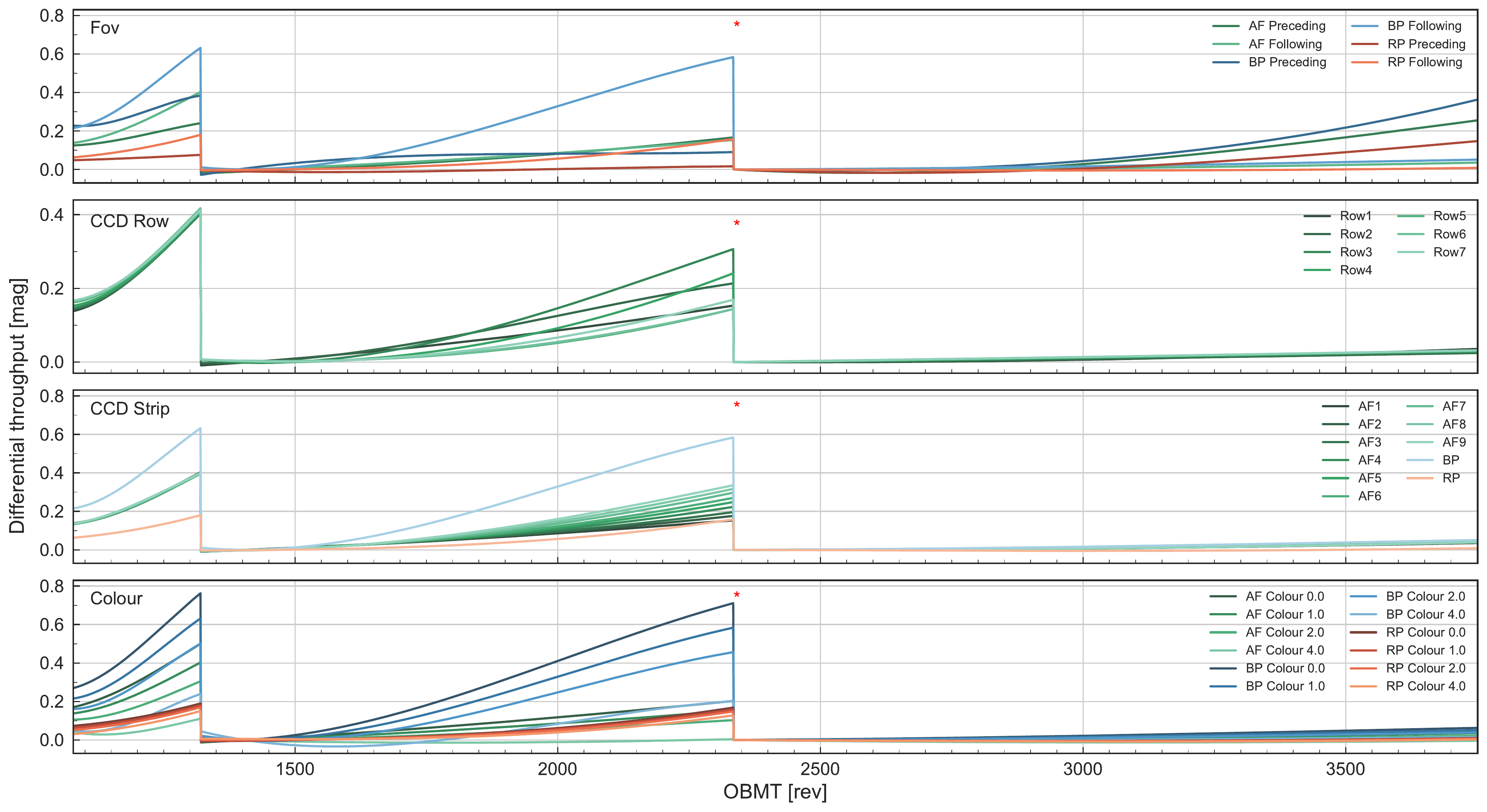}
\caption{
  Differential throughput $\Delta\tau$ (see \equref{tlc}) w.r.t. 2335 rev as a function of time,
  the reference epoch is indicated by the red asterisk.
  \emph{First Panel}: Variation in throughput for the two FoV in CCD row 1 for AF1 (green), BP
  (blue), and RP (red) showing larger contamination in the following FoV at least up to the second
  decontamination.
  \emph{Second Panel}: Variation in throughput for AF1 and following FoV for different CCD rows
  showing stronger contamination in the bottom part of the focal plane (low CCD row number). The range
  of the ordinate axis in this panel is more compressed since it does not show BP and RP.
  \emph{Third Panel}: Variation in throughput in the following FoV for all CCDs in row 1 showing an
  increase in contamination from AF1 to AF9; the maximum effect is in BP and the lowest effect
  in RP, as expected. \emph{Fourth Panel}: Variation in throughput in the following FoV, CCD row 1 for
  AF1, BP, and RP and different source colours showing that bluer sources are more heavily affected by
  contamination, hence the overall much larger systematic in the BP band.
  \label{fig:tlc}
}
\end{figure*}

In the early \PP\ test runs after the start of nominal operations (November 2014), it was discovered
that a time--dependent level of contamination was causing linear trends of $\approx0.0023$ mag/day
in the EPSL epoch photometry produced by \PP. This linear trend was caused by the varying level of
contamination that affected the data.
Contamination introduces a systematic offset in the bootstrap reference flux of a given source, which
is a function of the time distribution of the individual transits and the source colour (since the
size of the systematic effect caused by contamination on a given transit depends on the colour of the
source).
This systematic offset is imprinted on the reference fluxes and is not efficiently removed by the
iterative calibration loop described above. When solving for the various LS calibrations, this effect
causes over--/under-- corrections resulting in the linear trend reported in the test campaign.
With additional iterations, the linear trend was reduced to $\approx0.0021$, $\approx0.0018,$ and
$\approx0.0016$ mag/day, thus showing a rather slow decrease in the size of the effect.
In order to mitigate for this systematic effect without requiring a large number of iterations, we
introduced a new calibration that tracked the differential contamination level as a function of time. The
model fits the magnitude differences of epochs of a given source as a function of time and source
colour using a cubic time dependence and a linear colour--time cross term, as explained in more detail
below.

To measure the throughput loss due to contamination, we could compare the observed raw flux of sources
to their known true flux from other catalogues. However, to avoid introducing uncertainties and
systematic effects due to passband differences and colour transformations, we preferred not to use any
external catalogue. We instead devised a method to recover the variation in throughput with an arbitrary
constant defining the real throughput at a given time. This method allowed us to recover the
throughput evolution using only \gaia\ data.

We assume the throughput loss, in magnitiudes, to be a function of time $t$ and source colour $C$ and
express it using Chebyshev polynomials $T_n$ as basis functions, allowing for cross--terms between
time and colour:

\begin{equation}
  \tau(t, C) = \sum{a_nT_n(t)} + \sum{b_mT_m(C)} + \sum{c_mT_m(t)T_m(C)}
.\end{equation}
We can thus express the observed variation in throughput between two observations of the same source
$k$ as the difference between the throughput function $\tau$ evaluated at the two times $t_i$ and
$t_j$,

\begin{equation}
  \Delta\tau(t_i, t_j, C_k) = \tau(t_i, C_k) - \tau(t_j, C_k)
.\end{equation}

From the definition above, it is clear that the $\Delta\tau(t_i, t_j, C)$ polynomial does not have a
zero point or a linear colour term as the corresponding terms cancel out. For the throughput function
$\tau,$ we consider a cubic time dependence $(n=0,\dots,3)$, a linear colour, and time--colour
dependence $(m=1)$.
Since during the period of interest there are two decontaminations, the full time-line can be modelled
piece--wise by having a $\Delta\tau$ function for each of the three time ranges plus a discontinuity in
time and colour at each decontamination event: $K_n(t_i, t_j, C) = d_o + d_1T_1(C)$. This leads us to
the following formulation:

\begin{eqnarray}
  \Delta\tau(t_i, t_j, C_k) &=& \Delta\tau_0(t_i, t_j, C_k) + K_1(t_i, t_j, C_k) +\label{eq:tlc}\\
  & & \Delta\tau_1(t_i, t_j, C_k) + K_2(t_i, t_j, C_k) +\nonumber\\
  & &\Delta\tau_2(t_i, t_j, C_k).\nonumber
\end{eqnarray}

When producing the least-squares solution for \equref{tlc}, a pair of observations of a given source
$k$ only contributes to the $\Delta\tau$ polynomials containing one of the two observations, and it
will only activate the discontinuity function $K_i$ if the two observations are separated by the $i$-th
decontamination. Given a set of observations of a source, there are many different ways to form pairs
of observations: we found that a good coverage of the time range span by the observations is created
by interleaving the $N$ observations: $(z_i, z_{N/2+i})$, where $i=0,\dots,N/2$.

To solve for the differential throughput function, we need sources with observations providing a good
coverage of the time-line: the simplest way to achieve this is to select in each HEALPix pixel
\citep[of level six, see][]{HPIX} the $N$ sources with the most observations (we used $N=20$).
To limit the total number of sources to a manageable level, we selected sources in the (uncalibrated)
magnitude range $G=[13.0, 13.5]$ and introduced a colour restriction to the (uncalibrated) range
$\xpcol=[0.0, 4.0]$ to normalise the colour for use with the Chebyshev basis.

\afigref{tlc} shows the variation in contamination with respect to a reference epoch of 2335 rev
(i.e. shortly after the end of the second decontamination, as indicated by the red asterisk) since it
seems safe to assume that the overall contamination is at its lowest absolute level at this time.
For all top three panels, we considered a source colour of $\xpcol=1.0$. The first panel (top) shows
that the contamination level is stronger in the following FoV, at least up to the second
decontamination (higher throughput loss), and that it is much stronger in BP than in RP, as expected from
the wavelength dependence of the contamination. The second panel shows that the contamination is
generally stronger at the bottom of the focal plane (lower row number, see \figref{fpa}) using the
AF1 CCD and the following FoV calibrations. The third panel shows that the contamination increases along
CCD row 1, and the effect is stronger in AF9 for the \gband. The fourth (bottom) panel shows the
colour--dependence of the throughput in the following FoV for CCD row 1 in AF1, BP, and RP using a
source colour $\xpcol=0.0, 1.0, 2.0, \text{and } 4.0$.

In \pdr, the introduction of this new link calibration reduced the linear trend in the EPSL to 0.00008
mag/day, but it did not completely remove it. The reason probably was that
although the model provides a reasonable approximation, it is not sophisticated enough to reproduce
all the systematic effects caused by contamination. This is especially true for the EPSL period, when
the contamination level was both most intense and showed the strongest time-variation (contamination was higher
during the commissioning phase, but this paper is only concerned with the observations obtained during science
operations). Section~\ref{sec:calplan} discusses our improved mitigation of contamination for \gdr.

\subsection{Gate window-class link calibration}\label{sec:gwlc}

%
%
%
%
%
%
%
The calibration process is complicated by the multitude of instrumental configurations with which
observations can be acquired. Each configuration is effectively a different instrument, and for
a given time range, the \PP\ system therefore produces a set of calibrations, one for each instrumental
configuration. For simplicity, we call these configurations calibration units (CU). For the LS, a CU
is identified by the FoV, the CCD row and strip (i.e. a given CCD), the active gate, and the window class. This leads
to a total of 2108 CUs for each time interval. As mentioned in \secref{instrument}, it is possible
that some transits are acquired with a non--nominal configuration when a higher priority simultaneous
transit triggers a gate activation. A total of 1848 possible non--nominal configurations
exist, which means that in a given time interval, there will be at most 3956 individual LS calibration
solutions.

%
%
%
%
%
%
%
For the SS, a CU is identified by the FoV, the CCD row and strip, the active gate, and a 4-pixel-wide AC bin.
Instead of explicitly modelling the high-frequency spatial variations of the AC CCD response (e.g. due
to bad/hot columns), we map the AC 1D flat field by computing an independent solution in equally sized
bins of 4 pixels. This leads to a total of 1,120,416 nominal CUs per time interval, which increases to
2,332,704 CUs when all the possible non--nominal configurations
are considered.

The LS and SS solutions are derived independently for each CU. The fact that sources are observed
multiple times in different configurations ensures that CUs are linked together (since all solutions
are computed using the same set of reference source fluxes), and therefore, the internal photometric
system is homogeneous over the entire instrument. Unfortunately, this is not true for all CUs. Owing to
the combination of narrow gate-activation magnitude ranges, the small number of bright sources available
and small uncertainties in the onboard magnitude detection, there is insufficient mixing between
some CUs. The iterative process used to establish the photometric system should be able to take care
of this, but the convergence could be very slow. To speed this process up, an additional link
calibration has been introduced for \gband\ and \xp\ integrated photometry. This calibration provides
the link between the different window--class and gate configurations and is applied only at stage 1 in
the calibration process (see \secref{calsum}) when the initial set of raw reference fluxes
that are used to bootstrap the iterative calibration process described above is derived. The links are computed from
multiple observations of the same source in different configurations.

In \pdr, a single set of calibrations was computed using $\approx10$ rev, and this was then used to calibrate
the entire dataset. The results were not optimal, as revealed by the features in the errors of the
final source photometry \citep[see \secname~7 in][]{DWE_DR1}.
For \gdr,\ we computed a set of calibrations for each week using $\approx8$ consecutive revs, therefore
calibrating possible time variations for this effect.

\subsection{DR2 calibration strategy}\label{sec:calplan}

In \pdr,\ the calibration process described in \secref{calsum} was applied to the entire dataset: this
was necessary to ensure a sufficient number of good-quality sources to establish the internal
photometric system. Since \gdr\ spans nearly two years, it provides a much better sky coverage than
DR1. This allowed us to be more selective in which data to use for the initialisation of the
photometric system. In particular, it is clear from Fig.~\ref{fig:tlc} that in the period after the
second decontamination, the throughput loss is much lower and more stable. This period spans
$\approx354$ days and therefore provides nearly two complete sky coverages. We therefore decided to
use this subset of data, which we call the \calinit\ dataset/period, to initialise the
photometric system following the procedure outlined in \secref{calsum}. We refer to the subset
from the beginning of operations up to the second decontamination as the \calonly\ dataset/period; the
motivation for the name is described below.

\begin{figure}
\begin{center}
\colfig{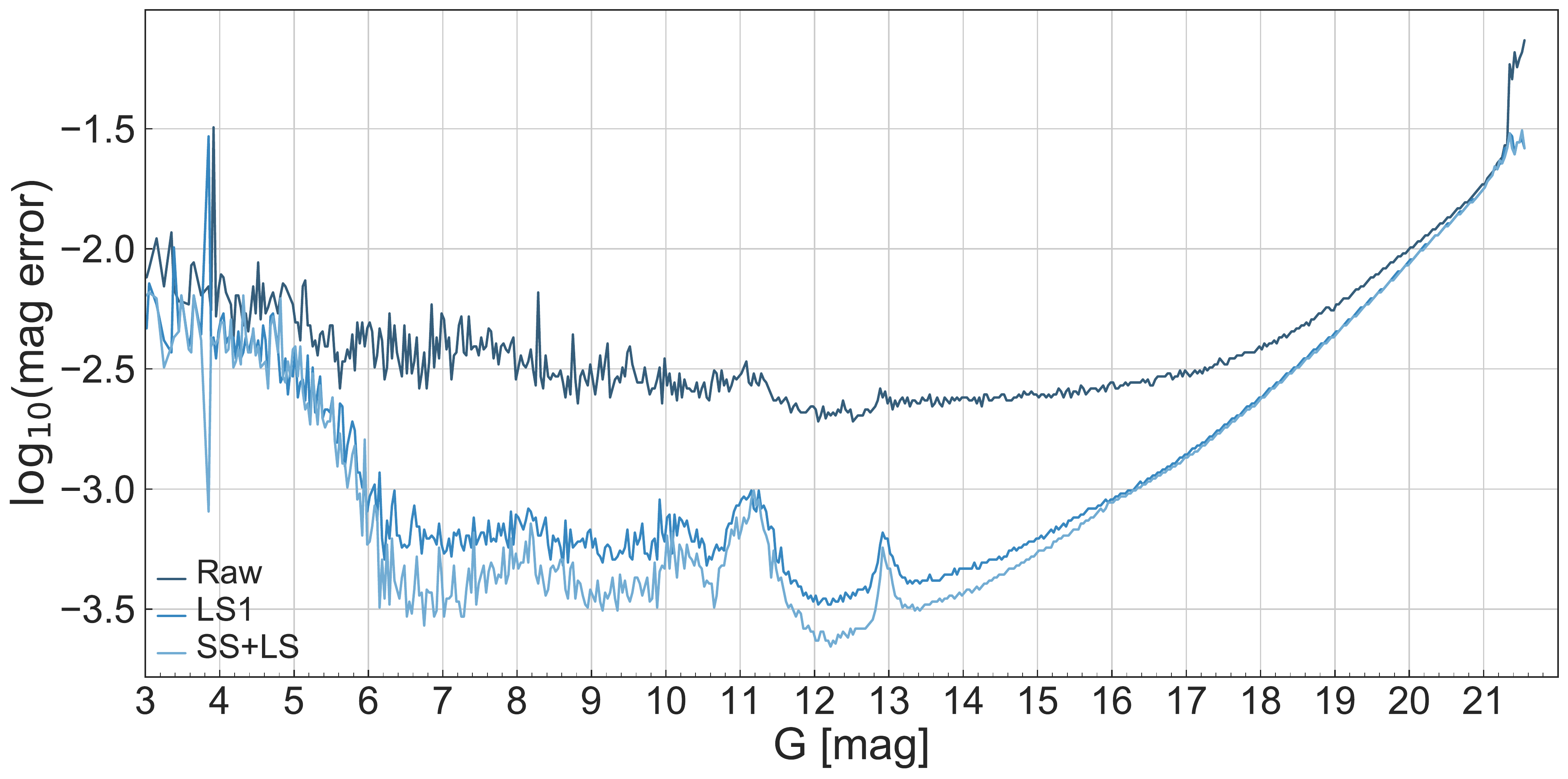}
\caption{
  Mode of the error distribution on the weighted mean \gband\ magnitude as a function of $G$
  magnitude for (1) the uncalibrated source photometry (Raw), (2) the source photometry
  obtained after the first LS solution in the calibration intialisation loop described in
  \secref{calsum} (LS1), and (3) the source photometry obtained after the last SS+LS iteration.
  The increased scatter at $G\leq13$ is related to changes in the observation configuration
  (e.g. window class/gates), the limitations in the IPD algorithms, and the handling of
  saturation and flux loss. A more in--depth discussion can be found in \cite{DWE_DR1,DWE_DR2}.
  \label{fig:photcal}
}
\end{center}
\end{figure}

Using all observations in the \dsinit,\ we generated the uncalibrated source photometry from which we
selected the sources to be used to compute the time--link calibration. Although the effect of
contamination is much reduced in the \dsinit, the variations in throughput are non--negligible, and
therefore it is still appropriate to perform the time--link calibration.
The main difference with respect to the model described in \secref{tlc} is that only a single period
is required as there are no discontinuities to take into account.
The time--link calibrations were then applied to the entire \dsinit\ to generate a new set of
reference source fluxes that were used to solve for the gate window--class link calibration as
described in \secref{gwlc}. Finally, we computed a new set of reference source fluxes by applying
both the time--link and gate window--class link calibrations to all applicable observations in the
\dsinit. This provides an improved set of reference fluxes to bootstrap the internal photometric
calibration, as described in \secref{calsum} and shown in \figref{calloop}.

The next stage is the calibration loop, in which we iteratively solve for the LS calibration and then
produce an updated set of reference source fluxes to be used in the subsequent iteration. We performed
a total of five iterations and refer to the process as the `LS iterations'.
The final stage of the initialisation of the internal photometric system involves the introduction of
the SS calibration \citep[see \secref{calsum} and also \secname~4 in][]{JMC_DR1}. In this stage, we
iterate between the LS and SS calibrations without updating the reference source fluxes. An iteration
is composed of two steps: in the first, we solve for the SS calibration using LS--calibrated
observations; in the second, we solve for the LS calibration using SS--calibrated observations based
on the SS calibrations obtained in the first step. We performed two of these iterations. The final
internal photometric system is then established by generating a new set of reference source fluxes by
applying the last set of SS and LS calibrations. A single set of SS calibrations (composed of
1,749,013 independent solutions) was computed using the entire \dsinit: we had indeed already
confirmed in \pdr\ that the SS calibration is very stable and
has no significant variations over a timescale of one year \citep{DWE_DR1, DWE_DR2}.

\afigref{photcal} shows the mode of the error distribution on the weighted mean \gband\
magnitude as a function of $G$ magnitude derived using the \dsinit\ for the source photometry
generated from uncalibrated observations, observations calibrated using the first LS solution from the
initialisation loop (see above and \secref{calsum}), and the observations calibrated using the final
set of SS and LS solutions. As expected, the introduction of the LS
calibration is a great improvement, but further improvement due to the subsequent iterations in the initialisation loop and
the introduction of the SS calibration is also quite noticeable, especially at $G<16$. The large
scatter at $G<6$ is mainly due to saturation, whereas the various bumps are caused by a combination
of changes in the instrumental configuration (window class, gate) and the limitations in the PSF/LSF
models used in \gdr. We refer to \cite{DWE_DR1, DWE_DR2} for a detailed analysis
of the error properties of the \gaia\ photometry and a discussion of the various features in the error
distributions.

\begin{figure*}
\centering
\widefig{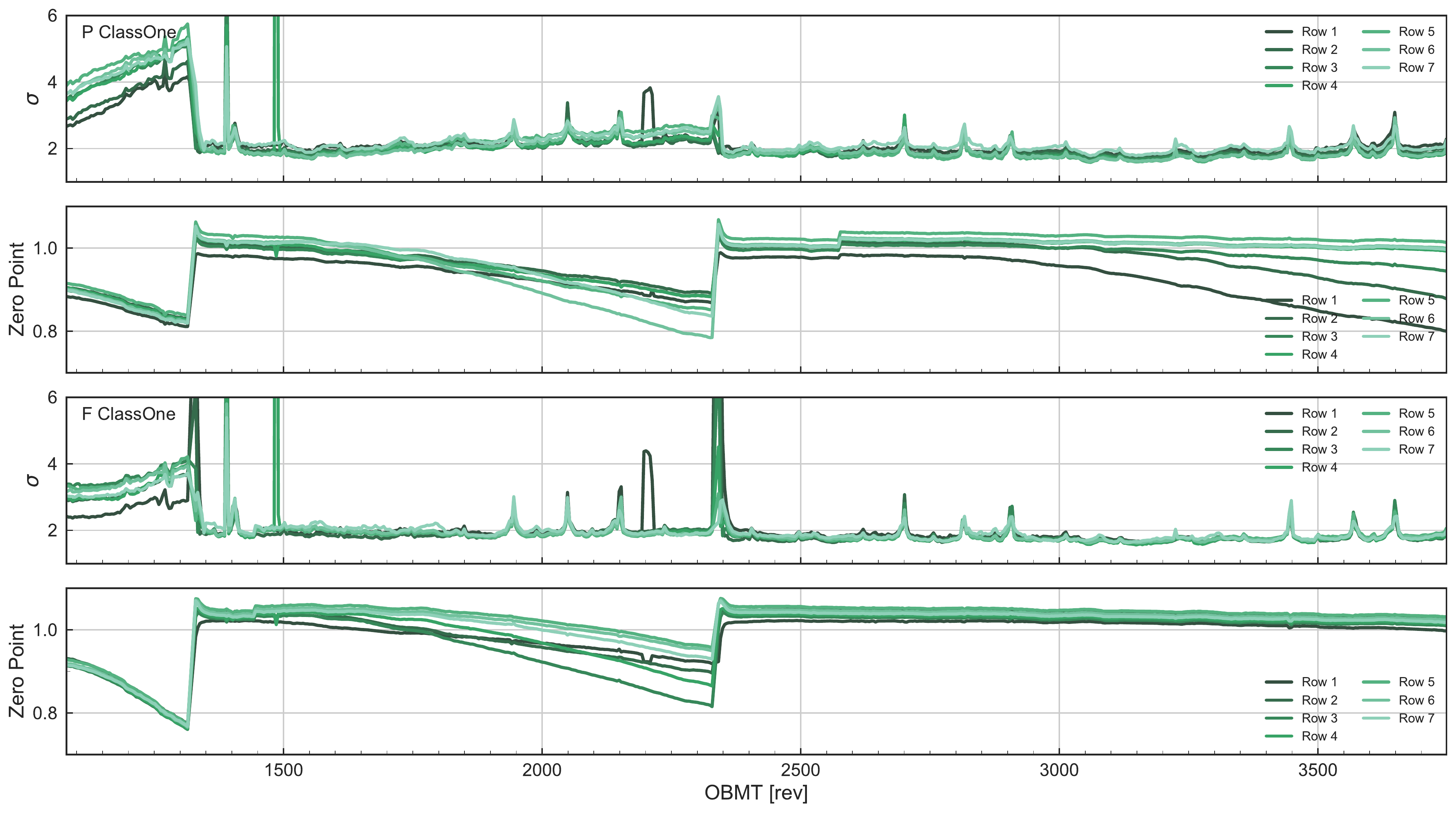}
\caption{
  Time evolution of the LS calibration standard deviation (sigma) and zero point for the preceding FoV
  (panels 1 and 2, respectively) and following FoV (panels 3 and 4, respectively) for each CCD row.
  All solutions only consider the ungated AF1 observations acquired with window class 1
  (corresponding approximately to $13<G<16$). The two decontaminations are clearly visible in both
  FoVs as a major discontinuity in the calibration zero point. The first refocus can be seen as a
  very small step in the zero point for the following FoV, whereas the second refocus is more visible
  as a slightly larger step in the zero point for the preceding FoV. Additional features visible in
  both the standard deviation and zero point are discussed in the text.
  \label{fig:lscals}}
\end{figure*}

The set of source reference fluxes generated from the \dsinit\ can now be used to produce the LS and
SS calibrations for the \dsco. This involves two SS--LS iterations in the same fashion as for the
\dsinit: a single set of SS calibrations was also derived for this period.
We then perform one final SS and LS calibration on the \dsinit\ to have a consistent set of SS and LS
calibrations for both the \calinit\ and \calonly\ datasets based on the same photometric system. The
linear trend in the EPSL (see \secref{tlc}) period caused by varying contamination has now been
further reduced from 0.00008 mag/day of DR1 to 0.000015 mag/day in \gdr: this amounts to 0.4
milli-magnitiudes over the 28 days of EPSL.

\afigref{lscals} provides an example of the time evolution of the standard deviation and zero point
of the final LS calibrations. The most striking features are the decrease in the overall standard
deviation `floor' after each decontamination event and the remarkable agreement between the
calibration zero-point time evolution in the CCD rows and FoV and what was independently measured via
the time--link calibration (see \secref{tlc} and \figref{tlc}). We also note that in the period
leading to the first decontamination, the standard deviation of the solutions
progressivily deteriorated as the contamination built up. This trend is only marginally visible in the
preceding FoV (top panel) in the central period between first and second decontamination. This is
explained by the fact that the PSF/LSF models were generated using 15 rev in the period after the
second decontamination and therefore become progressively worse at representing the data when the
contamination level increases and varies.
In the period after the second decontamination, the following FoV calibrations are extremely stable,
whereas in the preceding FoV, it is possible to see a significant correction at the time of the
second refocus ($\approx2750$ rev) and the more pronounced throughput loss in the lower rows caused
by the increase in contamination level in the period.
Occasional $\delta$-functions such as spikes in the standard deviation are due to individual calibration
solutions that are affected by an anomalously large number of poor observations that are mainly caused by sub-optimal
calibrations (e.g. background) used in the IPD process and are not a cause of major concern since
they are naturally taken care of by the DPAC iterative processing.

\subsection{Source photometry}\label{sec:srcphot}

The set of LS and SS calibrations for the \calonly\ and \calinit\ periods and the reference source
SSCs can now be used to produce the final source photometry. We note that the error distribution of
the individual transits of a given source is, in general, heteroscedastic since the observations are
taken under a variety of different instrumental configurations. We therefore generate the source
photomety as the weighted mean of the individual calibrated observations using the inverse variance
as the weight \citep[see \secname~6 in][for more details]{JMC_DR1}. The source photometry is produced
by applying the SS and LS calibrations to all individual transits of a source followed by the
computation of the weighted mean from all calibrated transits. For the $G$ band we included only the
AF CCDs since the SM is always observed with the Gate12 configuration. This means that saturation and
low photon counts will always be a problem at the bright and faint end, respectively. Moreover, the SM
observations are obtained in 2D windows with a sampling such that the effective pixels are twice the
size of a standard AF CCD. 

When validating the source photometry published in \pdr, we discovered that in some cases, it was
highly affected by outliers. In particular, because the cross--match and spurious
detection black--listing process was still sub-optimal, transits from different sources could
occasionally be assigned by the cross--match to the same source.
If the magnitude difference between the sources was significant, the epochs from the fainter sources
would bias the weighted mean towards the faint end since their associated weights would be much
lower than for the brighter epochs. Occasional poor IPD results or the use of poor photometric
calibrations could also lead to similar results. For \gdr, the robustness of the source photometry
determination is improved by introducing a rejection process based on median statistics. We first
determine the median and median absolute deviation (MAD) of all valid calibrated observations (in a
given band) and then reject all observations that are more than $5\sigma$ from the median
(the standard deviation was obtained as $\sigma=1.4826$ MAD).
An observation is considered valid if it has been both SS and LS calibrated and if the calibrated
flux is higher than 1 $e^-/s$ ($G\approx26$); this is a very generous lower limit for a physically
meaningful flux.

In order to calibrate a transit of a given source, it is necessary to have the reference SSCs for the
source and the reference integrated BP and RP fluxes, which are required in the normalisation of the
SSC fluxes (see \appref{sscs}). When deriving the link--calibrated source photometry for the
bootstrapping of the photometric system initialisation loop, the time--link calibration could only be
applied to sources within the colour range used by the model ($\xpcol=[0.0, 4.0]$). All epochs of
bluer and redder sources could therefore not be calibrated in the standard procedure and
were excluded from the calibration process altogether.

In \gdr, we used three different approaches to generate the source photometry, which depends on the
availability of colour information for the source. We call the three procedures and the corresponding
samples of sources \textup{\textit{\emph{gold}, \emph{silver,}}} and \emph{\textit{\textup{bronze}}}.

\subsubsection{Gold sources}

We define as \textup{\textit{\emph{gold} }}any source for which the photometry was produced by the full calibration
process described in \secref{calplan}.
\PP\ produced a total of 1,527,436,167 gold sources. The actual number in the \gdr\ archive will probably
be lower because various data-quality filters are applied during the catalogue preparation
\citep[see][for more detail]{CU9}.
Sources were excluded from the gold sample mainly by the colour selection introduced by the time--link
calibration. However a number of sources that were originally within the time--link calibration colour
range dropped out of the gold sample during the iterative calibration process described in
\secref{calplan}. These dropouts are only a small fraction of the initial sample, and the main cause is
probably related to a small number of \xp\ transits that fail to be calibrated at some stage during
the iterations (see also \appref{sscs}).

\subsubsection{Silver sources}

To recover sources that were excluded from the gold sample, we implemented an iterative calibration
process that uses the SS and LS calibration produced from the \calonly\ and \calinit\ datasets to
update the mean source photometry starting from the uncalibrated mean source photometry. The effect
of the iterations is to produce progressively better source photometry by making use of improved
mean source colour information (SSCs). In \gdr, we define as \textit{\emph{silver}} any source that went
through this iterative calibration process: \PP\ produced a total of 144,944,018 silver
sources.
The actual number in the \gdr\ archive will probably be lower because various data-quality
filters are applied during the catalogue preparation \citep[see][for more detail]{CU9}.
Sources with incomplete reference source colour information (see \appref{sscs}) could not be
calibrated using this iterative process and therefore are not part of the silver sample. As we
noted for the gold sample, a small fraction of sources that were originally part of the silver
sample (i.e. at the first iteration) dropped out of the sample during the iterative calibration
process: the same conclusions as drawn for the gold sources apply.

\subsubsection{Bronze sources}

Transits for the remaining set of sources in principle are not calibratable since they miss the colour
information required to apply the LS calibrations. As a compromise between quality of the photometry
and completeness of the \gdr\ catalogue, we calibrated the remaining sample of sources using a set
of default SSC colours. These default colours were obtained from a subset of sources by converting
the individual source SSC fluxes into colours, then taking the median value of each SSC colour, and
finally renormalising these median SSCs to ensure that their sum is equal to one (see \appref{sscs}). \PP\
produced a total of 901,338,610 bronze sources, of which 861,630,440 have available
\gband\ photometry, 194,652,181 have integrated BP photometry, and 226,114,046 have integrated RP
photometry. The actual number in the \gdr\ archive will probably
be lower because various data-quality filters are applied during the catalogue preparation
\citep[see][for more detail]{CU9}. For all bronze sources available in the \gdr\ archive, only the \gband\ photometry is
published. We note that several of the bronze sources are likely to have extreme colours (which
would explain why so many are missing either BP or RP). Since $\approx44\%$ of the bronze sources have
$G>21$, it is likely that a significant fraction of these sources are not real and are caused instead
by spurious detections.

\begin{figure}
\begin{center}
\colfig{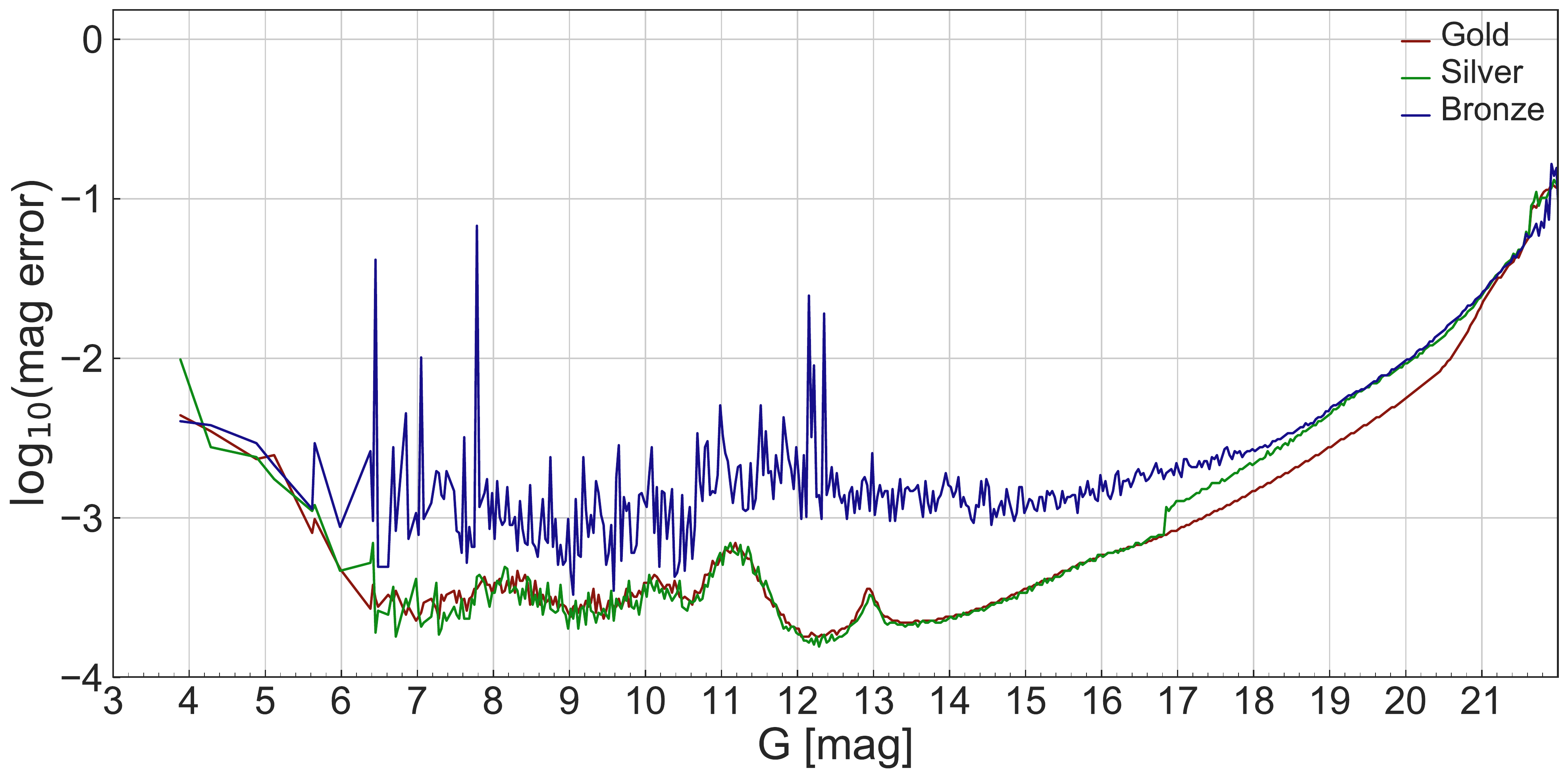}
\caption{
  Mode of the error distribution on the weighted mean \gband\ magnitude as a function of $G$
  magnitude for the gold (red), silver (green), and bronze sources (blue).
  \label{fig:srcflavours}
}
\end{center}
\end{figure}

\afigref{srcflavours} shows a comparison of the mode of the magnitude error distribution versus magnitude
for the $G$ band for gold, silver, and bronze sources. There is very good agreement
between the gold and silver source photometry errors: the two samples are indistinguishable up
to $G\approx16.8,$ where the error distribution mode is clearly discontinuous because towards the faint end,
the error distribution of the silver sources appears
to be bimodal. The reasons for this are not yet clear. It is possible that the bimodality is caused by
selection effects in the population of sources that end up in the silver calibration mode combined (or
not) with processing problems (e.g. background underestimation or crowding effects in \xp).
These effects are not visible in the bronze sample because the \xp\ information is not used at
all. The bronze photometry has considerably larger errors and scatter, as shown by the noise on the
mode line. We refer to \cite{DWE_DR2} for a more detailed discussions of the scientific quality of the
source photometry.

\section{\PP\ implementation details}\label{sec:pp}

The data-processing platform adopted for PhotPipe is the open--source Hadoop distributed processing
system. Hadoop is a mature system with wide adoption in industry for a variety of data processing
executed on large datasets.
Hadoop has been designed to operate well with commodity hardware and is composed of a distributed
file system (HDFS) that provides good resilience against hardware and network failure and against data loss
by means of data replication. The other core component is an application resource management layer
that allows the scheduling of distributed applications running on the cluster.

The version of the \PP\ processing system used for \gdr\ is entirely based on the Map/Reduce
programming model \citep{MapReduce}. The Map/Reduce paradigm is a very simple parallelisation model
that involves a data transformation stage (Map), a sorting and grouping by some user-defined key, and
a final transformation of the values associated to a given key (Reduce). The Hadoop implementation
distributes the processing tasks optimally by scheduling them on the node that holds a local copy of
the data. This approach provides both horizontally scalable I/O and processing capacity.
In \secref{mapred} we briefly recall the key concepts of the distributed Map/Reduce framework
\citep[see][for more details]{MapReduce}, and then we describe in the following section the
implementation of the iterative initialisation of the photometric system described in \secref{calsum}.

\subsection{Distributed Map/Reduce overview}\label{sec:mapred}

Given an input stream of key/value pairs of type $\{a,x\}$, the Map/Reduce model involves applying a
Map function transforming the $i$-th input key/value pair into a sequence of $n$ key/value pairs of
$\{b,y\}$
$$\textrm{map}: \{a_i, x_i\}\rightarrow(\{b,y\}_n).$$
Let the output key $b$ have $K$ distinct values, the output of the Map function is grouped into $K$
sets composed of the $k$-th $b$ key value and the sequence of values associated to that key. The
Reduce function is then applied to each of these sets transforming the input sequence $(y_j)$ keyed by
$b_k$ into a sequence of $p$ key/value pairs of type $\{c,z\}$,
$$\textrm{reduce}: \{b_k,(y_j)\}\rightarrow(\{c,z\}_p).$$
This simple model is implemented by Hadoop in a distributed fashion. The input dataset is stored in
the distributed file system, the data are segmented into blocks of equal size, each block is stored on a
node of the cluster, and the system ensures that there are always $R$ copies of any given data block stored
on $R$ different nodes (where $R$ is configurable per file, based on requirements of performance and
robustness against node failure). Assuming that the input data are composed of $B$ blocks, Hadoop will
schedule $B$ parallel Map tasks, each one applying the user-defined map function to all key/value pairs
in the assigned block. Hadoop attempts to schedule each of the tasks on a node that holds a copy of
the block to maximise I/O performance. If a map task fails (e.g. because of hardware/network
glitches or outages), Hadoop will automatically re-schedule the task on a different node; if all nodes
holding a copy of the input block are busy, Hadoop will schedule the task to another node and the input data
will be transferred over the network.
The parallelisation of the map stage is determined by the number of blocks in the input data and the
the cluster size (i.e. how many nodes and how many tasks per node can be executed).

The size of the
input dataset for the Reduce stage is unknown at scheduling time, so that the parallelisation is defined by
specifying the number $P$ of partitions in which the dataset set should be subdivided. Each map
output key/value pair is assigned by Hadoop to one of the $P$ partitions using a partitioning
function: each partition is assigned to a single reduce task.
The next stage is called shuffle and involves collecting all records belonging to a given partition
on the node that has been assigned the task of processing that partition. This stage involves fetching
the data over the network from multiple nodes. Each reducer process then merge-sorts the input data, groups
them by key, and applies the user-defined= reduce function to each set $\{b_k, (y)_j\}$. The $\{c,z\}_p$
outputs of each partition are then written back to the distributed filesystem.
The merge-sort process is very efficient since the outputs of each individual Map task are also
merge-sorted and therefore each individual Reduce node need only do one final merge-sort of the partial
Map outputs.
The default behaviour is for Hadoop to use hash-based partitioning and lexicographic byte
order for the sorting and grouping, but each one of these phases can be customised by supplying
a user-defined function: this feature is heavily used in the implementation of the photometric
calibration workflow.
In the following section we present an example of how this simple model involving the
definition of a map function, a reduce function, and, optionally, a sorting function and a grouping
function can be used to implement the calibration workflow described in \secref{calib} in \PP.

\subsection{Distributed LS initialisation}\label{sec:lsmapred}

\begin{figure}
\begin{center}
\colfig{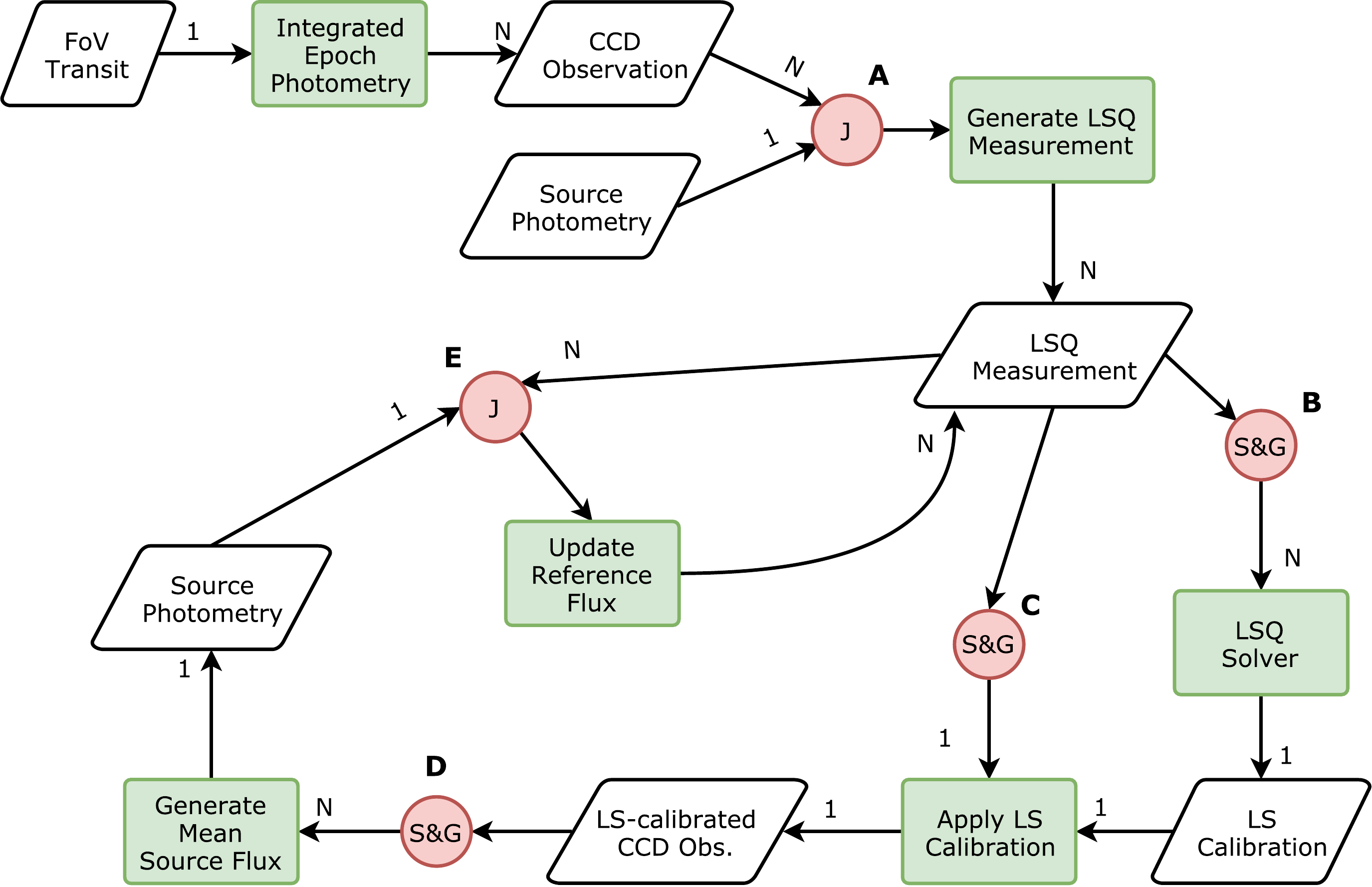}
\caption{
  Map/Reduce workflow for the initialisation of the photometric system (LS iterations, see
  \secref{calsum}). The workflow is composed of five Map/Reduce jobs labelled \textbf{A}
  to \textbf{E}. The red circles represent the global distributed join (\textbf{J}) or sorting and
  grouping by key (\textbf{S\&G}) operations. The green boxes represent processes: the input(s) are
  represented by the incoming arrow, and the output by the outgoing arrow. The data cardinality is
  shown on the labels as 1 when a single record is handled at a time, or N when multiple records
  are handled simultaneously. See the text for further information on the distributed workflow.
  \label{fig:lsmapred}
}
\end{center}
\end{figure}

The Map/Reduce implementation of the LS calibration iteration loop (see \figref{calloop} and
\secref{calsum}) is shown in \figref{lsmapred}. The workflow is composed by five jobs: the one--off
bootstrap job \textbf{A,} and the four jobs that comprise an LS iteration, \textbf{B} to \textbf{E}.

The first stage of the LS iterations is executed only once and involves generating the individual
CCD observations, including integrated \xp\ and SSCs, and attaching to each one the appropriate
reference flux information (from the source). This bootstrap job \textbf{A} consumes two input
streams: (1) the uncalibrated FoV transits that are converted into the integrated epoch photometry
(composed of the IPD \gband\ fluxes, integrated \xp\ fluxes, and SSC) and separated into the
individual CCD components keyed by the source identifier; and
(2) the reference source photometry records,
which are simply read and output keyed by source identifier.
At the reduce stage, all records associated with a given source identifier are collected and processed
by a single call of the reduce function, which will attach the appropriate reference flux and SSC
information to each indivudal CCD transit. We call the output type an LSQ measurement since it
represents an individual contribution to one of the LSQ problems producing a given LS calibration
solution.

The first job, \textbf{B}, of the calibration loop produces the set of LS calibration solutions, one
solution is produced per calibration unit and per time--range. This can be easily implemented by
assigning each input LSQ measurement to the corresponding time range and calibration unit: a single
reduce call will then receive all the LSQ measurements that contribute to a single LS calibration.
In this approach, however, the order in which the measurements are processed at the reduce stage is
not deterministic since it will depend on the order of completion of the various map tasks. Round--off
errors in the LSQ solution could then produce slightly different LS calibrations. In order to make the
LS solution deterministic (i.e. fully reproducible), we adopt a compound key containing the time range,
CU, and transit identifier.
We then use a custom sorting function that orders the LSQ measurements by calibration unit (using the
default lexicographic order), and for a given CU, by increasing transit identifier. We then provide
a custom grouping function that will perform the grouping based only on the CU and time-range ignoring
the transit identifier.

The next stage in the calibration loop involves calibrating the CCD observations, using the set of LS
calibrations produced by job \textbf{B}, to then generate a new version of the mean source photometry.
In principle, we could implement this as a single Map/Reduce job since the LS calibration application
process is performed on individual LSQ measurements (and hence could be taken care of in the Map
stage), and the source photometry requires all calibrated observations for a given source that can be
grouped by outputting the calibrated CCD observations keyed by source identifier at the Map stage.
The overall performance (total execution time) of a Map/Reduce job depends heavily on its concurrency:
everything else being equal, the more Map (and later Reduce) tasks that can be executed simultaneously
on the cluster, the faster the job will complete. The maximum number of concurrent tasks that can run
on a single cluster node is limited by the amount of memory available; to maximise
concurrency, it is therefore important to minimise the memory footprint of the individual tasks. In our specific
case, the input LSQ measurements to be calibrated are not time ordered
(because job \textbf{A} involves a join by source identifier and hence the output of a given reduce
  task will be ordered by lexicographic byte of the source identifier hash value) this means that a given
Map task would need to keep in memory the entire set of LS calibrations
(amounting to several gigabytes). This process can be made more memory efficient by ordering the input
LSQ measurements in time: since a Map task only processes a subset of records, it would thus be
necessary to keep in memory only a small subset of the LS calibration (several megabytes).
For this reason, we perform the generation of the new source photometry in two Map/Reduce jobs.
Job \textbf{C} reads the LSQ measurements and outputs them keyed by transit identifier: the LSQ
measurements reach the reducer sorted by time (since the 42 most significant bits of the transit
  identifier represent the acquisition time of the AF1 CCD of that transit).
At the Reduce stage, we only need to keep a limited number of calibrations in memory since the input
data are time ordered. The reducer outputs calibrated CCD observations. Since the source photometry is composed of
several passbands ($G$, \gbp, \grp\ , and the eight \xp\ SSCs) we define a key
containing the source identifier, the transit identifier, and the CCD/SSC information.
Job \textbf{D} then reads these calibrated CCD observations and outputs them keyed by the compound
key defined above. The job uses a custom sorting function that orders the CCD observations
for a given source by increasing transit identifier (i.e. increasing time) and increasing CCD/SSC.
The job also uses a custom grouping function that will only consider the source identifier component,
thus collecting all CCD observations for each source in the order specified above.
In this way, the reducer can efficiently compute the source photometry for each band by simply
accumulating (see \secref{accum}) the input calibrated fluxes until a change in CCD/SSC is detected:
at that point, the mean flux for this band is finalised and the computation for the following band
started.
Finally, job \textbf{E} closes the calibration loop by updating the original LSQ measurements with
the new reference source fluxes produced by job \textbf{D}. This is a simple join that supplies each
call of the reduce function with all LSQ measurements and the mean photometry (in all bands) for a
given source.

\subsection{Performance}\label{sec:perf}

The processing required for the generation of the \gdr\ calibrated photometry involved a variety of
Map/Reduce jobs with different properties: some were I/O--bound, some were CPU--bound, some involved
memory-intensive operations, and others had to perform demanding join operations on hundreds of
billions of records. A detailed analysis of the performance properties considerering all these factors
is clearly beyond the scope of this paper. However, since this paper presents the first use of the Hadoop
and Map/Reduce algorithms in a large-scale astrophysical survey, we believe it is still appropriate to
provide some general performance figures that highlight how successful the choice of this processing
platform has been for the \gaia\ photometric data.

Since we have covered the Map/Reduce implementation of the LS iteration loop in
\secref{lsmapred} in more detail, we consider the overall performance of this processing sequence in the
context of the initialisation of the photometric system (see \secref{calsum}). The performance metrics
for the five LS iterations are listed in \tabref{perfig}. Each iteration is composed of jobs from
\textbf{B} to \textbf{E}, as described in the previous section, and the jobs were run on the
Cambridge Hadoop cluster (see \appref{hardware} for more information). Overall, the 20 jobs completed
in 2.8 days, corresponding to nearly 21.5 CPU years. The reduce stage dominates the run time, while the
map stage only accounts for $\approx38\%$. This is expected since for all these jobs, the map stage
is essentially just reading data from HDFS and applying trivial transformations (e.g. key generation).
On the other hand, the reduce stage is responsible both for the LSQ solutions (with iterative
rejection) producing the LS calibration and for the computation of the mean source photometry. The job
duration is not equal to the sum of the Map and Reduce stage durations since the reducer tasks are
normally started before the completion of the map stage. This allows starting the transfer of the outputs
of the completed Map tasks to the nodes that will reduce the corresponding partition, hence reducing the
delay in starting the actual Reduce stage when the Map stage is completed.

\tabref{perfig} shows that the vast majority of the Map tasks have been reading from a local disk:
this is indeed crucial in allowing the distributed \PP\ processing to scale with the input data
volume. The read throughput is higher than expected, averaging $\approx3$ GB/sec, because the
operating system can use some of the memory for caching to further optimise the I/O.
Finally, we note that the Map stage experiences a very low number, $\approx0.04\%$, of task failures
caused by hardware and network glitches, which is normal when processing hundreds of terabytes of data
on a system of this scale (see \appref{hardware}).

\begin{table}
\begin{center}
\begin{tabular}{lrrr}
\hline
Metric & Map & Reduce & Total\\
\hline
Wall clock time & 26.19 hr & 61.22 hr & 68.21 hr\\
CPU time & 3194.28 hr & 4650.61 hr & 7844.89 hr\\
Number of tasks & 296,047 & 46,005 & 342,052\\
Data local tasks$^\dagger$ & $96.2\%$ & -- & --\\
Failed tasks & 141 & 1 & 142\\
Records I/O$^\ast$ & $7.74\times10^{12}$ & $3.78\times10^{12}$ & --\\
HDFS I/O$^\ast$ & 274.77 TB & 132.42 TB & --\\
\hline
\end{tabular}
\end{center}
\caption{
  Cumulative performance metrics for the 20 Map/Reduce jobs required to run five LS iterations for the
  initialisation of the photometric system. A single iteration is composed of job \textbf{B} to
  \textbf{E,} as described in \secref{mapred}. See the text for the discussion. $\dagger$ The concept
  of a data-local task is only meaningful for the Map stage. $\ast$ When reporting I/O figures, the
  Map stage reports the input figure (since the data are read off the distributed file system), and the
  Reduce stage reports the output figure (since the job results are written to the distributed
  file system).
  \label{tab:perfig}
}
\end{table}

\section{Concluding remarks and future developments}\label{sec:theend}

Producing science--grade data products from the \gaia\ raw data poses several challenges that are due to the
intrinsic complexity of the payload and acquisition system, the huge data volume and granularity, and
the necessity of a self--calibration approach. There are simply no full-sky surveys with the same spatial
resolution, high accuracy, and precision that \textit{Gaia} could use for the purpose of
photometric calibration.
In this paper we presented how these challenges were successfully overcome to design and implement
a distributed photometric processing system, \PP, which was used to produce the \gdr\ source photometry
in $G$ band and \xp.

The software architecture, design, and implementation have proved to be very stable during the entire
processing phase. Hadoop has proven to be an excellent choice for the core processing architecture
with zero downtime due to hardware/system problems.

A significant portion of the overall processing
time was dedicated to the validation of the photometric calibration process. These validation tasks
have also been implemented as map/reduce jobs: with nearly two billion sources (and two orders of
magnitude more epochs), visual inspection is clearly not an option. We therefore took the approach
of generating the distributions of various key metrics to be able to quickly assess the quality and
progress of the processing. Several examples have been shown in \cite{DWE_DR1} for \pdr\ and
more are available in \cite{DWE_DR2} for \gdr.

Although the pre--processing stages that were run for \gdr\ are reliable and able to mitigate the
instrumental effects they deal with, the background mitigation in \xp\ requires further improvement to
better handle the cases in which the astrophysical background dominates the straylight
contribution. This mostly affects the faint sources (e.g. $G>18$) where the local background becomes a
significant fraction of the overall flux. Another improvement that will be introduced in future data
releases for \xp\ is related to the handling of crowding effects, which are not limited to the
faint end, but affect the full magnitude range and can mimic variability because epochs
are acquired with different scanning directions and different overlapping fields of view, as dictated
by the satellite scanning law.

Overall, the iterative initialisation of the photometric system performed very well and produced a
noticeably better system than in DR1 \citep[see][]{DWE_DR2}, not only thanks to the improvements in
the IPD and \gband\ pre--processing, but also because of the possibility of using only mission data
with an overall lower and more stable level of contamination. Although we expect the approach described
in this paper to lead to an even better source photometry when the PSF/LSF models used in the IPD
process will include time and colour dependencies, some improvements are planned in the calibration
process itself. In particular, the gate and window--class link calibration (see \secref{gwlc}) does not
yet fully remove the discontinuities between the different instrumental configurations (see e.g.
\figref{photcal} and also \citealt{DWE_DR2}). At the bright end ($G<13$), saturation and
flux--loss effects become important. In principle, both effects are handled by the IPD process, but only
for the full 2D windows (i.e. $G<11$). For the 1D windows, the IPD will handle saturation effects, but
not flux loss (due to the lack of AC resolution). Although flux
loss does not appear explicitly in the
calibration models used for \gdr, we should note that the current model provides already a
calibration for the average flux--loss experienced by a source. This is equivalent to the case where
the source is perfectly centred within the window and there is no AC motion of the source along the
CCD transit. The centring and AC motion are different for each epoch of a given source and essentially
depend on the scanning law and the random errors in the VPU detection process.

An analysis of the residuals of the epoch photometry error (defined as the difference between
  the predicted AC position of the source on the CCD, derived from the source astrometry and
  satellite reconstructed attitude, and the window centre)
shows that the centring error distribution of the epochs acquired with window class 1 has a standard
deviation of 0.44 pixels and is centred on zero. The residual distribution of the epochs with a
centring error in the $1\sigma$ range has a standard deviation of 0.005 mag. When considering this
second--order effect on the source photometry, the size of this systematic will be much smaller
(depending on the number of epochs).
Although this effect could in principle be included by the LS calibration model, we note that at the
faint end, the effect will be harder to measure reliably because other systematic effects become more
important (e.g. background problems and crowding effects due to unresolved sources).
In \gdr,\ we decided not to include terms providing the second-order correction of the flux--loss (due
to centring error and AC motion) because we expected the effect to be smaller than the overall
improvements introduced by the better IPD and calibration strategy. The choice of whether to
include these terms will have to be re--evaluted for the next data release: since the IPD process will
be different (including colour and time dependency in the PSF/LSF models), it is hard to establish
using the current data whether the centring error and AC motion effects will be the same as is seen in
the \gdr\ data.

Although we are very pleased with the overall performance achieved by \PP\ on the Cambridge cluster
(see \appref{hardware}) for this data release, more work is required to ensure that \PP\ is
able to perform equally well in future data releases. One major challenge is posed by the fact that
the Map/Reduce allows only one global distributed operation (i.e. sorting+grouping/joining):
this means that when the algorithms involve more than one distributed operation, the implementation
requires the chaining of several Map/Reduce jobs, therefore generating a large amount of intermediate
data. Writing the intermediate data and then reading it back (in the next Map/Reduce job) will
progressively slow down the processing as the data volume increases.
An obvious way to keep the system scaling is to avoid the persistence of these
intermediate data products as much as possible. This can be achieved by rephrasing a given processing
flow in terms of a directed acyclic graph (DAG) instead of a linear concatenation of Map/Reduce stages
by using, for example, Apache Spark\footnote{\url{https://spark.apache.org/}}.
Although the approach is different, it can be easily implemented without requiring a complete rewrite
of the existing software because Spark adopts a functional model for the definition of the dataflow
DAG. This allows one to `re--wire' the existing modules defining the various Map and Reduce stages using
the Spark API. We have performed extensive testing of this approach that has confirmed the benefits in
terms of performance.

\begin{acknowledgements} 
  This work presents results from the European Space Agency (ESA) space mission \gaia. \gaia\ data are
  being processed by the \gaia\ Data Processing and Analysis Consortium (DPAC). Funding for the DPAC
  is provided by national institutions, in particular the institutions participating in the \gaia\
  MultiLateral Agreement (MLA). The \gaia\ mission website is \url{https://www.cosmos.esa.int/gaia}.
  The \gaia\ Archive website is \url{http://gea.esac.esa.int/archive/}.
  
%
  This work has been supported by the United Kingdom Rutherford Appleton Laboratory, the United
  Kingdom Science and Technology Facilities Council (STFC) through grant ST/L006553/1, and the United
  Kingdom Space Agency (UKSA) through grant ST/N000641/1.

%
  This work was supported by the MINECO (Spanish Ministry of Economy) through grant ESP2016-80079-C2-1-R
  (MINECO/FEDER, UE) and ESP2014-55996-C2-1-R (MINECO/FEDER, UE) and MDM-2014-0369 of ICCUB
  (Unidad de Excelencia “Mar\'ia de Maeztu”).
%
  We also thank the Agenzia Spaziale Italiana (ASI) through grants I/037/08/0, I/058/10/0, 2014-025-R.0,
  and 2014-025-R.1.2015 to INAF, and the Italian Istituto Nazionale di Astrofisica (INAF).

   We thank the referee, Mike Bessell, for suggestions that helped improve this paper.
\end{acknowledgements}

%
%

\bibliographystyle{aa} 
\bibliography{refs} 

\begin{thebibliography}{20}
\expandafter\ifx\csname natexlab\endcsname\relax\def\natexlab#1{#1}\fi
\expandafter\ifx\csname url\endcsname\relax
  \def\url#1{{\tt #1}}\fi
\expandafter\ifx\csname urlprefix\endcsname\relax\def\urlprefix{URL }\fi

\bibitem[{{Arenou} \& {CU9}(2018)}]{CU9}
{Arenou} F., {CU9} T., 2018, \aap, this volume

\bibitem[{Bird(2010)}]{FunAlgo}
Bird R., 2010, Pearls of Functional Algorithm Design, Cambridge University
  Press, New York, NY, USA, 1st edn.

\bibitem[{{Boyadjian}(2008)}]{PLM00108}
{Boyadjian} J., 2008, EADS Astrium/ESA

\bibitem[{{Carrasco} et~al.(2016){Carrasco}, {Evans}, {Montegriffo}
  et~al.}]{JMC_DR1}
{Carrasco} J.M., {Evans} D.W., {Montegriffo} P., et~al., Nov. 2016, \aap, 595,
  A7

\bibitem[{{Casta\~neda} et~al.(2018){Casta\~neda}, {Clotet}, {Gonz\'alez-Vidal}
  et~al.}]{IDU_XM}
{Casta\~neda} J., {Clotet} M., {Gonz\'alez-Vidal} J.J., et~al., 2018, \aap,
  this volume

\bibitem[{{de Bruijne} et~al.(2015){de Bruijne}, {Allen}, {Azaz}
  et~al.}]{VpuDetection}
{de Bruijne} J.H.J., {Allen} M., {Azaz} S., et~al., Apr. 2015, \aap, 576, A74

\bibitem[{Dean \& Ghemawat(2008)}]{MapReduce}
Dean J., Ghemawat S., Jan. 2008, Commun. ACM, 51, 107,
  \urlprefix\url{http://doi.acm.org/10.1145/1327452.1327492}

\bibitem[{{Evans} et~al.(2017){Evans}, {Riello}, {De Angeli} et~al.}]{DWE_DR1}
{Evans} D.W., {Riello} M., {De Angeli} F., et~al., Apr. 2017, \aap, 600, A51

\bibitem[{{Evans} et~al.(2018){Evans}, {Riello}, {De Angeli} et~al.}]{DWE_DR2}
{Evans} D.W., {Riello} M., {De Angeli} F., et~al., 2018, \aap, this volume

\bibitem[{{Fabricius} et~al.(2016){Fabricius}, {Bastian}, {Portell}
  et~al.}]{IdtRef}
{Fabricius} C., {Bastian} U., {Portell} J., et~al., Nov. 2016, \aap, 595, A3

\bibitem[{{Gaia Collaboration} et~al.(2016{\natexlab{a}}){Gaia Collaboration},
  {Brown}, {Vallenari} et~al.}]{GDR1}
{Gaia Collaboration}, {Brown} A.G.A., {Vallenari} A., et~al., Nov.
  2016{\natexlab{a}}, \aap, 595, A2

\bibitem[{{Gaia Collaboration} et~al.(2016{\natexlab{b}}){Gaia Collaboration},
  {Prusti}, {de Bruijne} et~al.}]{GaiaRef}
{Gaia Collaboration}, {Prusti} T., {de Bruijne} J.H.J., et~al., Nov.
  2016{\natexlab{b}}, \aap, 595, A1

\bibitem[{{G{\'o}rski} et~al.(2005){G{\'o}rski}, {Hivon}, {Banday}
  et~al.}]{HPIX}
{G{\'o}rski} K.M., {Hivon} E., {Banday} A.J., et~al., Apr. 2005, \apj, 622, 759

\bibitem[{{Hambly} et~al.(2018){Hambly}, {Cropper}, {Boudreault} et~al.}]{BIAS}
{Hambly} N.C., {Cropper} M., {Boudreault} S., et~al., 2018, \aap, this volume

\bibitem[{{Lindegren} et~al.(2012){Lindegren}, {Lammers}, {Hobbs}
  et~al.}]{AGIS_REF}
{Lindegren} L., {Lammers} U., {Hobbs} D., et~al., Feb. 2012, \aap, 538, A78

\bibitem[{{Lindegren} et~al.(2016){Lindegren}, {Lammers}, {Bastian}
  et~al.}]{TGAS}
{Lindegren} L., {Lammers} U., {Bastian} U., et~al., Nov. 2016, \aap, 595, A4

\bibitem[{{Lindegren} et~al.(2018){Lindegren}, {Hern{\'a}ndez}, {Bombrun}
  et~al.}]{AGIS_DR2}
{Lindegren} L., {Hern{\'a}ndez} J., {Bombrun} A., et~al., 2018, \aap, this
  volume

\bibitem[{{Pancino} et~al.(2012){Pancino}, {Altavilla}, {Marinoni}
  et~al.}]{SPSS}
{Pancino} E., {Altavilla} G., {Marinoni} S., et~al., Nov. 2012, \mnras, 426,
  1767

\bibitem[{{van Leeuwen}(1997)}]{FVL_HIPREV}
{van Leeuwen} F., Aug. 1997, \ssr, 81, 201

\bibitem[{White(2012)}]{Hadoop}
White T., 2012, Hadoop: The Definitive Guide, O'Reilly Media, Inc.

\end{thebibliography}

\vfill
\appendix

\section{Spectral shape coefficients}\label{sec:sscs}

The SSCs and their use in the photometric calibrations to provide colour information are described in
\secname~4 and 5 of \cite{JMC_DR1}. In this section we briefly recall the key concepts to clarify the
discussion of the source photometry grade (gold, silver, bronze) in \secref{srcphot}.

\begin{figure}[h!t]
\begin{center}
\colfig{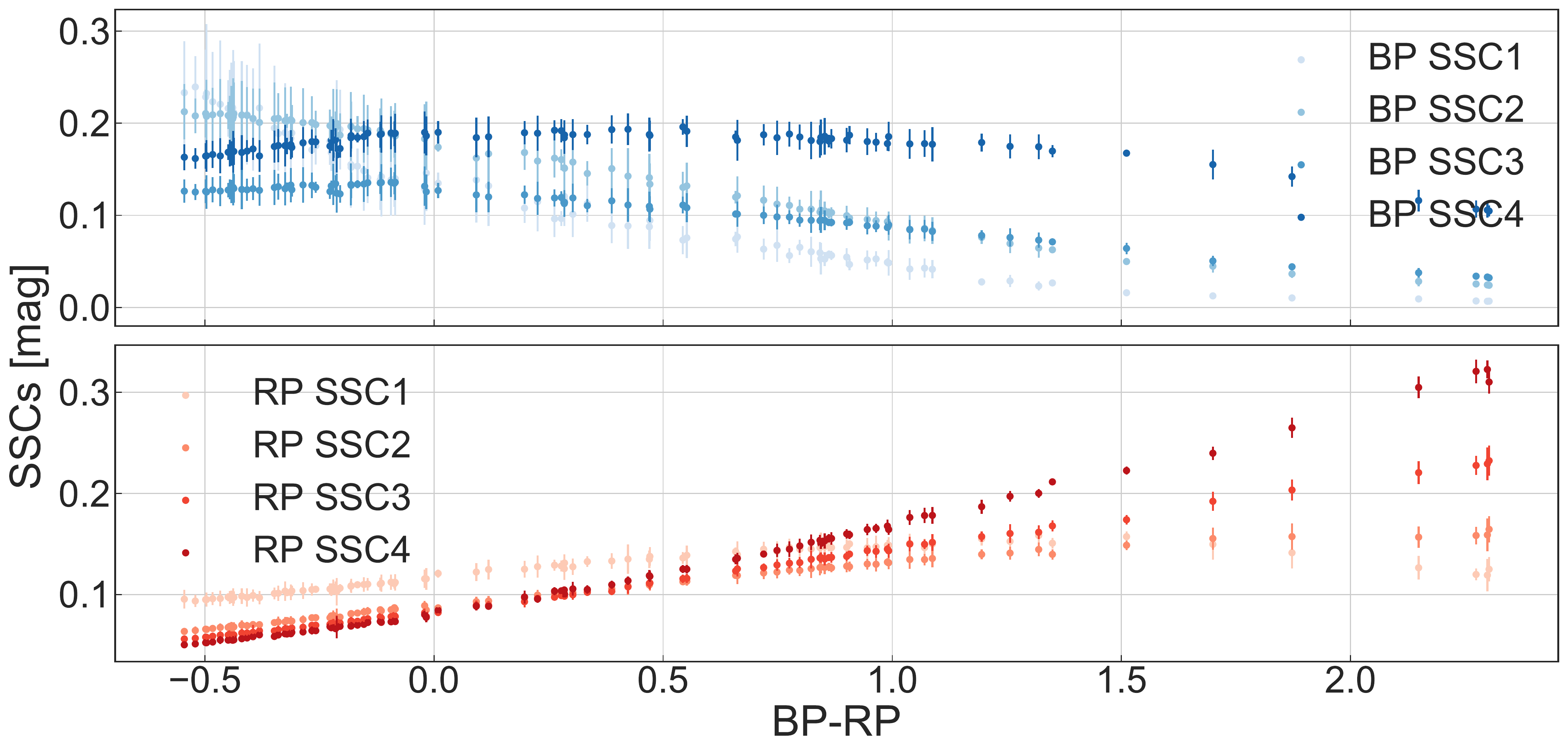}
\caption{
  Distribution of the 8 SSCs derived from the BP (top) and RP (bottom) spectra for the set of sources used in the
  external calibration process.
  \label{fig:sscs}}
\end{center}
\end{figure}

For both BP and RP, we defined four rectangular bands and produced the integrated flux in each band
for each transit: the results of this synthetic photometry are four BP SSCs and four RP SSCs per transit.
All eight SSCs are independently calibrated in the same fashion as the $G$ band and integrated \xp. The
calibrated epoch SSCs for each source are then used to compute the weighted-average source SSCs
fluxes.
These calibrated source SSC fluxes are then used in the LS calibration model to provide colour
information. The four BP SSCs are normalised so that their sum is equal to one, and the same is done for RP.
These colour SSCs are then used in the BP and RP LS calibration models. For the
calibration of the $G$ band instead both the BP and RP SSCs are used. In this case we apply an
additional normalisation to the colour SSCs such that their total sum is equal to one and the ratio
of the sum of the BP and RP colour SSCs is equal to the ratio of the integrated BP and RP fluxes.
\afigref{sscs} shows the distribution of the source SSCs for the SPSS used in the external
calibration. In this case the same normalisation as was used for the \gband\ calibration was applied.

\begin{figure}[h!t]
\begin{center}
\colfig{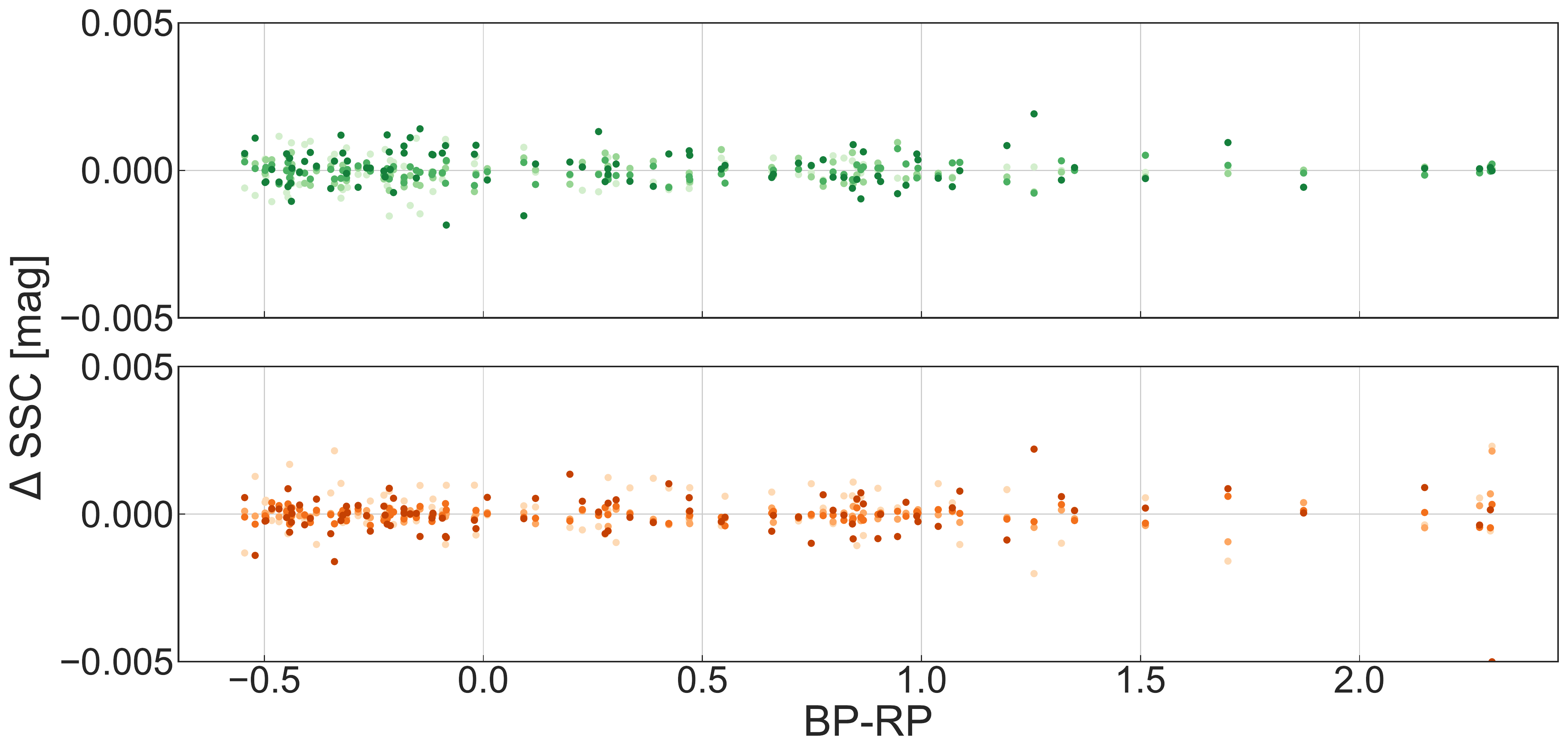}
\caption{
  Comparison between two sets of SSCs: the first is computed using the final geometric
  calibrations generated for \gdr, and the second is computed adding random noise of the same
  level as the scatter in the geometric calibration. The top panel shows the result for the BP SSCs,
  and the bottom panel shows the comparison for the RP SSCs.
  \label{fig:sscDiff}}
\end{center}
\end{figure}

In \secref{geocal} we described that the level of uncertainty in the geometric calibration does
not significantly affect the computation of the source SSCs. To confirm this, we have compared the set
of source SSCs shown in \figref{sscs} with an alternative set of SSCs computed by applying a geometric
calibration with an additional random noise of the same level as the scatter of the geometric
calibration solution in a period with no discontinuities.
The difference between the two sets of source SSCs is shown in \figref{sscDiff}, where the range
covered by the plots is equivalent to a few milli-magnitudes. There is no evidence for systematic
differences with colour, and the overall scatter is well within the uncertainties of the SSCs
themselves.

One important requirement of using source mean SSCs in the calibration model is that for any given
source
\begin{itemize}
\item
all four BP SSC average source fluxes must be available in order to apply the LS calibration solution
to an epoch BP flux to produce the internally calibrated flux (and analogously for RP),

\item
all eight SSC average source fluxes and the integrated \xp\ average source fluxes must be available in
order to apply the LS calibration solution to a \gband\ epoch flux to produce the internally
calibrated flux.
\end{itemize}
These requirements can become problematic, especially at the faint end and for sources with more
extreme colours because the synthetic photometry of the epoch spectra might fail to produce a valid
flux for one or more of the SSC bands. If this happens systematically for all transits of a given
source, then it will not be possible to calibrate these transits since the source colour information
(as represented by the eight SSC fluxes) might be incomplete or missing altogether.

\section{Weighted-mean source photometry by accumulation}\label{sec:accum}

For efficiency, \PP\ implements the computation of the weighted-mean flux in a given band for a given
source as a left-fold operation \citep[e.g.][]{FunAlgo} on the sequence of calibrated observations.
This is implemented by adding the contribution of an individual calibrated observation to three
accumulators that can then be used to generate the weighted-mean flux and error, $\chi^2$, variance
and scatter measures, and an estimate of the additional scatter caused by variability
(see \secname~6 in \citealt{JMC_DR1} and Eq. 87 of \citealt{FVL_HIPREV} ). For a given
passband, the fold operation is based on three accumulators and the total number of contributing
observations, $N$:
\begin{eqnarray}
A_1&=&\sum{w_i}\\
A_2&=&\sum{f_iw_i}\\
A_3&=&\sum{f_i^2w_i}
,\end{eqnarray}
where $f_i$ represents the flux of the $i$-th observation, and $w_i=1/\sigma_i^2$ is the associated
weight defined as the inverse variance.

\section{Hadoop cluster}\label{sec:hardware}

The Hadoop cluster used for the \gdr\ photometric data processing is hosted by the High Performance
Computing Service of the Univeristy of Cambridge, UK. The cluster is composed of 218 identical nodes
that serve both as storage nodes (i.e. contributing to the Hadoop distributed file system) and as
compute nodes (i.e. to run the distributed \PP\ processing jobs). Each node features dual 12 core Intel
E5-2650v4 2.2 GHz processors, 256 GB RAM, a 64GB system SSD, and 6 2TB 7.2k SAS hard drives.
Overall, the cluster provides a raw capacity of 2.32 PB of storage, 54.5 TB RAM, and 5232 physical cores that are
configured to run with hyperthreading to yield 10464 virtual cores. All nodes are interconnected using three
different networks: 1) 56Gb infiniband with a simple tree-topology for high-performance transfers for
the Map/Reduce job and data loading; 2) a standard 1Gb ethernet network for system monitoring and backup; and
3) a service network for remote system management. All cluster servers are running on the Scientific Linux 7.2
operating system.

\section{Acronyms}\label{sec:acro}

Below, we list the acronyms we used in this paper.\hfill\\
\begin{tabular}{lp{0.35\textwidth}}\hline\hline 
\textbf{Acronym} & \textbf{Description} \\\hline
%
AC&ACross scan (direction) \\\hline
AF&Astrometric field (in Astro) \\\hline
AGIS&Astrometric global iterative solution \\\hline
AL&ALong scan (direction) \\\hline
API&Application programming interface \\\hline
BP&Blue photometer \\\hline
CCD&Charge-coupled device \\\hline
DPAC&Data processing and analysis consortium \\\hline
DR1&(\gaia) data release 1 \\\hline
DR2&(\gaia) data release 2 \\\hline
DR3&(\gaia) data release 3 \\\hline
EPSL&Ecliptic Pole scanning law \\\hline
FoV&Field of view \\\hline
FWHM&Full width at half--maximum \\\hline
GP&Galactic Plane \\\hline
GPS&Galactic-Plane scan \\\hline
HDFS&Hadoop distributed file system \\\hline
IDT&Initial data treatment \\\hline
IDU&Intermediate data update \\\hline
IPD&Image parameter determination \\\hline
LS&Large scale \\\hline
LSF&Line spread function \\\hline
LSQ&Least squares \\\hline
MAD&Median absolute deviation (from the median)\\\hline
OBMT&On-board mission timeline \\\hline
PSF&Point--spread function \\\hline
RP&Red photometer \\\hline
SM&Sky mapper \\\hline
SS&Small scale \\\hline
SSC&Spectrum shape coefficient \\\hline
SPSS&Spectro--photometric standard star \\\hline
TCB&Barycentric coordinate time \\\hline
TDI&Time--delayed integration (CCD) \\\hline
TGAS&Tycho-Gaia astrometric solution \\\hline
UTC&Coordinated universal time \\\hline
VO&Virtual object \\\hline
VPU&Video processing unit \\\hline
\end{tabular}

\end{document}